\documentclass[peerreview]{IEEEtran}
\usepackage{cite,subfigure,psfrag,graphicx,amssymb,makeidx,multicol,footmisc,}
\usepackage[tbtags]{amsmath}
\usepackage[usenames,dvipsnames]{pstricks}
\usepackage{epsfig}
\DeclareMathOperator*{\diag}{diag} \DeclareMathOperator*{\bdiag}{bdiag}
\DeclareMathOperator*{\tr}{tr}

\DeclareMathOperator*{\rank}{rank}
\newtheorem{Theorem}{Theorem}

\include{MyBib.bib}
\begin{document}
\title{Optimal Multiuser Zero-Forcing with Per-Antenna Power Constraints for Network MIMO Coordination }
\author{\authorblockN{Saeed Kaviani and Witold A. Krzymie\'{n}}\\
\authorblockA{University of Alberta / TRLabs, Edmonton, Alberta, Canada T6G 2V4\\
E-mail: \{saeedk,wak\}@ece.ualberta.ca}%
\thanks{The work discussed in this paper was presented in part at the WCNC 2008, Las Vegas, NV, March-April 2008.}}
\maketitle
\begin{abstract}
 We consider a multi-cell multiple-input multiple-output (MIMO) coordinated downlink transmission, also known as network MIMO, under per-antenna power constraints. We investigate a simple multiuser zero-forcing (ZF) linear precoding technique known as block diagonalization (BD) for network MIMO. The optimal form of BD with per-antenna power constraints is proposed. It involves a novel approach of optimizing the precoding matrices over the entire null space of other users' transmissions.
 An iterative gradient descent method is derived by solving the dual of the throughput maximization problem, which finds the optimal precoding matrices globally and efficiently. The comprehensive simulations illustrate several network MIMO coordination advantages when the optimal BD scheme is used. Its achievable throughput is compared with the capacity region obtained through the recently established duality concept under per-antenna power constraints.
\end{abstract}

\section{Introduction}
While the potential capacity gains in point-to-point \cite{Foschini98,Telatar99} and multiuser \cite{Gesbert07} multiple input multiple output (MIMO) wireless systems are significant, in cellular networks this increase is very limited due to intra and
inter-cell interference. Indeed, the  capacity gains promised by MIMO are severely degraded in cellular environments \cite{Blum03,Dai04}. To mitigate this limitation and achieve spectral efficiency increase due to MIMO spatial multiplexing in future broadband cellular systems, a network-level interference management is necessary. Consequently, there has been a growing interest in network MIMO coordination \cite{Karakayali06c,Karakayali06d,Somekh07,ZhangDai04,Jing07,Gesbert10}. Network MIMO coordination is a very promising approach to increase signal to interference plus noise ratio (SINR) on downlinks of cellular networks without reducing the frequency reuse factor or traffic load. It is based on cooperative transmission by base stations in multiuser, multi-cell MIMO systems. The network MIMO coordinated transmission is often analyzed using a large virtual MIMO broadcast channel (BC) model with one base station and more antennas \cite{Shamai02,Goldsmith03,Jafar04}.
This approach increases the number of transmit antennas to each user and hence the capacity increases dramatically compared to conventional MIMO networks without coordination \cite{Karakayali06d,ZhangDai04,Somekh07}. Moreover, inter-cell scheduled transmission benefits from the increased multiuser diversity gain \cite{choi08}. The capacity region of network MIMO coordination as a MIMO BC has been previously established under sum power constraint \cite{Shamai03,Vishwanath03,Yu04,Weingarten06,Goldsmith03a} using uplink-downlink duality. However, the coordination between multiple base stations requires per-base station or even more realistic in practice per-antenna power constraints to be extendable to any linear power constraints. Under per-antenna power constraints, uplink-downlink duality for the multi-antenna downlink channel has been established in \cite{Yu06,Yu07} using Lagrangian duality framework in convex optimization \cite{Boyd04} to explore the capacity region. It is known that the capacity region is achievable with dirty paper coding (DPC). However, DPC is too complex for practical implementations. Consequently, due to their simplicity, linear precoding schemes such as multiuser zero-forcing or block diagonalization (BD) are considered \cite{Spencer04,Murch04}.


\par The key idea of BD is linear precoding of data in such a way that transmission for each user lies within the null space of other users' transmissions. Therefore, the interference to other users is eliminated. Multi-cell BD has been employed explicitly for network MIMO coordinated systems in \cite{Zhang09,karakayali07,Hadisusanto08,Liu09} with the diagonal structure of the precoders and the sum power constraint \cite{Spencer04}. Although there were attempts in these papers to optimize the precoders to satisfy per-base-station and per-antenna power constraints, this structure of the precoders is no longer optimal for such power constraints and must be revised \cite{karakayali07,Wiesel08,Huh09}. In \cite{Boccardi06}, the ZF matrix is confined to the pseudo-inverse of the channel for the single receive antenna users with per-antenna power constraints. The sub-optimality of pseudo-inverse ZF beamforming subject to per-antenna power constraints was first shown in \cite{karakayali07}. \cite{Wiesel08} presented the optimal precoders' structure using the concept of generalized inverses which lead to a non-convex optimization problem and the relaxed form requires semi-definite programming (SDP) \cite{Vandenberghe98}. This is investigated only for single-antenna mobile users. \cite{Huh09} also uses the generalized inverses for the single-antenna mobile users, but using a multistage optimization algorithms.
\par In this work, we aim to maximize the throughput of network MIMO coordination employing multiple antennas both at the base stations and the mobile users through optimization of precoding. An optimal form of BD is introduced by extending the search domain of precoding matrices to the entire null space of other users' transmissions \cite{Kaviani08}. The dual of throughput maximization problem is utilized to obtain a simple iterative gradient descent method \cite{Boyd04} to find the optimal linear precoding matrices efficiently and globally. The gradient descent method applied to the dual problem requires fewer optimization variables and less computation than comparable algorithms that have already been proposed in \cite{Huh09,Wiesel08,Zhang09,Hadisusanto08}. \cite{Zhang10} has employed the idea presented in \cite{Kaviani08} which is optimizing over the entire null space of other users's channels but a sub-gradient algorithm is obtained.  The sub-gradient method is not a descent method unlike the gradient method and does not use the line search for the step sizes \cite{Boyd03}. Furthermore, our approach is also extended to the case of non-square channel matrices, single-antenna mobile users and per-base-station power constraints. In contrast to previous numerical results on network MIMO coordination \cite{Karakayali06b,Zhang09,Huang09} assuming the sum power or per-base-station power constraints, in this paper the proposed optimal BD is examined with per-antenna power constraints enforced. To consider feasible network MIMO coordination in practice, local coordination of base stations is used through clustering \cite{Zhang09,Venkatesan07,Huang09}. The results show that the proposed optimal BD scheme outperforms the earlier BD schemes used in network MIMO coordination.  For the sake of comparison the capacity limits are determined employing the uplink-downlink duality idea in MIMO BC under per-antenna power constraint introduced in \cite{Yu06,Yu07}.
%
\par The remainder of this paper is organized as follows. In Section \ref{Sec:SystemModel} the system model is introduced, and the network MIMO coordination structure, the transmission strategy and the corresponding capacity region are discussed.  In Section \ref{Sec:Multi-CellBD} the multi-cell BD scheme is studied and its comparison with the conventional BD is presented, which motivates research on optimal multi-cell BD under per-antenna power constraints. The optimal multi-cell BD scheme is proposed in Section \ref{Sec:OptimalBD} and its further extensions and generalizations are considered. Comprehensive numerical results are presented in Section \ref{Sec:NumericalResults} following the discussion of the simulation setup in Section \ref{Sec:SimulationSetup}. Conclusions are given in Section \ref{Sec:Conclusions}.

\section{System Model}\label{Sec:SystemModel}
\subsection{Network MIMO Coordinated Structure}

We consider a downlink cellular MIMO network, with multiple antennas at both base stations and mobile users. Each user is equipped with $n_r$ receive antennas and each base station is equipped with $n_t$ transmit antennas. The base stations across the network are assumed to be coordinated via high-speed back-haul links. For a large cellular network of several cells, this coordination is difficult in practice and requires large amount of channel state information and user data available at each base station. Hence, clustering of the network is applied, where each group of $B$ cells is clustered together and benefits from intra-cluster coordinated transmission \cite{Venkatesan07,Huang09,Zhang09}. Hence, within each cluster each user's receive antennas may receive signal from all $N_t=n_tB$ transmit antennas. The cellular network contains $C$ clusters. The base stations within each cluster are connected and capable of cooperatively transmitting data to mobile users within the cluster. Hence, there are two types of interference in the network, the intra-cluster and inter-cluster interference. If we define $\mathbf H_{c,k,b} \in \mathbb C^{n_r \times n_t}$ to be the downlink channel matrix of user $k$ from base station $b$ within cluster $c$, then the aggregate downlink channel matrix of user $k$ within cluster $c$ is a $n_r \times N_t$ matrix defined as \mbox{$\mathbf H_{c,k}=\left[\mathbf H_{c,k,1} \mathbf H_{c,k,2} \cdots \mathbf H_{c,k,B}\right]$}. The aggregate downlink channel matrix for all $K$ users scheduled within cluster $c$, $\mathbf H_c \in \mathbb C^{Kn_r \times N_t}$ is defined as \mbox{$\mathbf H_c=[\mathbf H_{c,1}^\textsf{T} \cdots \mathbf H_{c,K}^\textsf{T}]^\textsf{T}$}, where $(\cdot)^\textsf{T}$ denotes the matrix transpose. The multiuser downlink channel is also called broadcast channel (BC) in information theory literature \cite{Cover91}. Assuming that the same channel is used on the uplink and downlink, the aggregate uplink channel matrix is $\mathbf H_c^\textsf{H}$, where $(\cdot)^\textsf{H}$ denotes the conjugate (Hermitian) matrix transpose \cite{Goldsmith03}. The multiuser uplink channel is also called multiple-access channel (MAC). In the BC, let $\mathbf x_c \in \mathbb C^{N_t \times 1}$ denote the transmitted signal vector (from $N_t$ base stations' antennas of $c$th cluster) and let $\mathbf y_{c,k} \in \mathbb C^{n_r\times 1}$ be the received signal at the receiver of the mobile user $k$. The noise at receiver $k$ is represented by $\mathbf n_{c,k} \in \mathbb C^{n_r \times 1}$ containing $n_r$ circularly symmetric complex Gaussian components ($\mathbf n_{c,k} \sim \mathcal{CN}(0,\sigma^2\mathbf I)$). The received signal at the $k$th user in cluster $c$ is then
\begin{equation}\label{ReceivedSignal}
\begin{split}
\mathbf y_{c,k}=&\underbrace{\mathbf H_{c,k}\mathbf x_{c}}_{\text{Intra-cluster signal}}+\underbrace{\sum_{\hat{c}=1, \hat{c} \neq c}^{C}
\mathbf H_{\hat{c},k} \mathbf x_{\hat{c}}}_{\text{Inter-cluster interference}}+ \underbrace{\mathbf n_{c,k}}_{\text{noise}}
\end{split}
\end{equation}
where $\mathbf H_{\hat{c},k}$ represents the channel coefficients from the surrounding clusters $\hat{c}$ to the $k$th user of the cluster $c$. The transmit covariance matrix can be defined as $\mathbf S_{c,\mathbf x} \triangleq \mathbb E\left[\mathbf x_c\mathbf x_c^\textsf{H}\right]$. The base stations are subject to the per-antenna power constraints $p_1, \ldots,p_{N_t}$, which imply
\begin{equation}\label{PAPC}
\left[\mathbf S_{c,\mathbf x}\right]_{ii} \leq p_i, \quad i=1,\ldots,N_t
\end{equation}
where $\left[\cdot\right]_{ii}$ is the $i$th diagonal element of a matrix.
\par The cancelation of intra-cluster multiuser interference is done by applying BD, which is discussed in Section \ref{Sec:Multi-CellBD}. The remaining inter-cluster interference plus noise covariance matrix at the $k$th user of the cluster $c$ is given by
 \begin{equation}
  \mathbf R_{c,k}=\mathbb E\left[\mathbf z_{c,k}\mathbf z_{c,k}^\textsf{H}\right]=\mathbf I_{n_r}+\sum_{\hat{c}=1, \hat{c}\neq c}^{C} \mathbf H_{\hat{c},k}\mathbf S_{\hat{c},\mathbf x} \mathbf H_{\hat{c},k}^\textsf{H}.
   \end{equation}
   where $\mathbb E\left[\mathbf x_{\hat{c}}\mathbf x_{\hat{c}}^\textsf{H}\right]=\mathbf S_{\hat{c},\mathbf x}$.

To simplify the analysis, we have normalized the vectors in (\ref{ReceivedSignal}) dividing each by the standard deviation of the additive noise component, $\sigma$. Completely removing the inter-cluster interference requires universal coordination between all surrounding clusters. The worst-case scenario for interference is when all surrounding clusters transmit at full allowed power \cite[Theorem 1]{Ye05}. Although this result is for the case with the total sum power constraint on the transmit antennas, it is used in our numerical results and it gives a pessimistic performance of the network MIMO coordination \cite{Huang09}. Then, a pre-whitening filter can be applied to the system and as a result the inter-cluster
 interference in this case can be assumed spatially white \cite{Shim08}. The received signal for the $k$th user in the $c$th cluster after post-processing can be simplified as
 \begin{equation}\label{ReceivedSignal1}
  \mathbf y_k=\mathbf H_k \mathbf x+\mathbf z_k, \quad k=1,\ldots,K.
  \end{equation}
  where $\mathbf z_k$ is the noise vector. For ease of notation, we dropped the cluster index $c$.
\subsection{Capacity Region for Network MIMO Coordination}\label{Sec:CapacityRegion}
The capacity region of a MIMO BC with sum power constraint has been previously discussed in
\cite{Shamai03,Vishwanath03,Yu04}.  The sum capacity of a Gaussian vector broadcast channel under per-antenna power constraint is the saddle-point of a minimax problem and it is shown to be equivalent to a dual MAC with linearly constrained noise \cite{Yu07}. The dual minimax problem is convex-concave and consequently the original downlink optimization problem can be solved globally in the dual domain. An efficient algorithm using Newton's method \cite{Boyd04} is used in \cite{Yu07} to solve the dual minimax problem; it finds an efficient search direction for the simultaneous maximization and minimization. This capacity result is used to determine the sum capacity of the multi-base coordinated network and it constitutes the performance limit for the proposed transmission schemes.
\subsection{Transmission Strategy}\label{Sec:TransmissionStrategy}

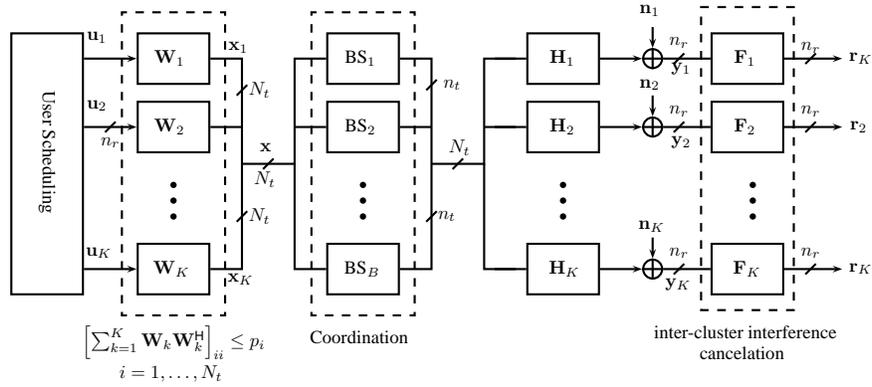
\begin{figure*}[!t]
\centering
\scalebox{.7} 
{
\begin{pspicture}(1,-3.658125)(19.242813,3.658125)
\psframe[linewidth=0.04,dimen=outer](5.7809377,3.0596876)(4.3809376,2.0596876)
\psframe[linewidth=0.04,dimen=outer](5.7809377,1.7596875)(4.3809376,0.7596875)
\psframe[linewidth=0.04,dimen=outer](5.7809377,-0.9403125)(4.3809376,-1.9403125)
\psline[linewidth=0.04cm,arrowsize=0.05291667cm 2.0,arrowlength=1.4,arrowinset=0.4]{->}(3.3809376,2.5596876)(4.3809376,2.5596876)
\psline[linewidth=0.04cm,arrowsize=0.05291667cm 2.0,arrowlength=1.4,arrowinset=0.4]{->}(3.3809376,1.2596875)(4.3809376,1.2596875)
\psline[linewidth=0.04cm,arrowsize=0.05291667cm 2.0,arrowlength=1.4,arrowinset=0.4]{->}(3.3809376,-1.4403125)(4.3809376,-1.4403125)
\psdots[dotsize=0.12](5.0809374,0.1596875)
\psdots[dotsize=0.12](5.0809374,-0.1403125)
\psdots[dotsize=0.12](5.0809374,-0.4403125)
\psline[linewidth=0.04](5.7809377,2.5596876)(6.3809376,2.5596876)(6.3809376,-1.4403125)(5.7809377,-1.4403125)(5.7809377,-1.4403125)
\psline[linewidth=0.04](5.7809377,1.2596875)(6.3809376,1.2596875)(6.3809376,1.2596875)(6.3809376,1.2596875)
\psline[linewidth=0.04](6.3809376,0.5596875)(7.3809376,0.5596875)(7.3809376,2.5596876)(7.9809375,2.5596876)
\psline[linewidth=0.04](7.3809376,0.5596875)(7.3809376,-1.4403125)(7.9809375,-1.4403125)
\psline[linewidth=0.04cm](7.9809375,1.2596875)(7.3809376,1.2596875)
\psframe[linewidth=0.04,dimen=outer](9.380938,3.0596876)(7.9809375,2.0596876)
\psframe[linewidth=0.04,dimen=outer](9.380938,1.7596875)(7.9809375,0.7596875)
\psframe[linewidth=0.04,dimen=outer](9.380938,-0.9403125)(7.9809375,-1.9403125)
\psline[linewidth=0.04](9.380938,2.5596876)(9.980938,2.5596876)(9.980938,-1.4403125)(9.380938,-1.4403125)
\psline[linewidth=0.04cm](9.380938,1.2596875)(9.980938,1.2596875)
\psline[linewidth=0.04](9.980938,0.5596875)(10.980938,0.5596875)(10.980938,2.5596876)(11.580937,2.5596876)
\psline[linewidth=0.04](10.980938,0.5596875)(10.980938,-1.4403125)(11.580937,-1.4403125)
\psline[linewidth=0.04cm](11.580937,1.2596875)(10.980938,1.2596875)
\psframe[linewidth=0.04,dimen=outer](16.680937,3.0596876)(15.280937,2.0596876)
\psframe[linewidth=0.04,dimen=outer](16.680937,1.7596875)(15.280937,0.7596875)
\psframe[linewidth=0.04,dimen=outer](16.680937,-0.9403125)(15.280937,-1.9403125)
\psline[linewidth=0.04cm](11.780937,1.2596875)(11.180938,1.2596875)
\psframe[linewidth=0.04,dimen=outer](13.180938,3.0596876)(11.780937,2.0596876)
\psframe[linewidth=0.04,dimen=outer](13.180938,1.7596875)(11.780937,0.7596875)
\psframe[linewidth=0.04,dimen=outer](13.180938,-0.9403125)(11.780937,-1.9403125)
\psline[linewidth=0.04cm](11.780937,2.5596876)(11.180938,2.5596876)
\psline[linewidth=0.04cm](11.780937,-1.4403125)(11.180938,-1.4403125)
\psline[linewidth=0.04cm,arrowsize=0.05291667cm 2.0,arrowlength=1.4,arrowinset=0.4]{->}(13.180938,2.5596876)(13.980938,2.5596876)
\psline[linewidth=0.04cm,arrowsize=0.05291667cm 2.0,arrowlength=1.4,arrowinset=0.4]{->}(13.180938,1.2596875)(13.980938,1.2596875)
\psline[linewidth=0.04cm,arrowsize=0.05291667cm 2.0,arrowlength=1.4,arrowinset=0.4]{->}(13.180938,-1.4403125)(13.980938,-1.4403125)
\pscircle[linewidth=0.04,dimen=outer](14.180938,2.5596876){0.2}
\psline[linewidth=0.04cm](13.980938,2.5596876)(14.380938,2.5596876)
\psline[linewidth=0.04cm](14.180938,2.7596874)(14.180938,2.3596876)
\pscircle[linewidth=0.04,dimen=outer](14.180938,1.2596875){0.2}
\psline[linewidth=0.04cm](13.980938,1.2596875)(14.380938,1.2596875)
\psline[linewidth=0.04cm](14.180938,1.4596875)(14.180938,1.0596875)
\pscircle[linewidth=0.04,dimen=outer](14.180938,-1.4403125){0.2}
\psline[linewidth=0.04cm](13.980938,-1.4403125)(14.380938,-1.4403125)
\psline[linewidth=0.04cm](14.180938,-1.2403125)(14.180938,-1.6403126)
\psline[linewidth=0.04cm,arrowsize=0.05291667cm 2.0,arrowlength=1.4,arrowinset=0.4]{->}(14.180938,3.1596875)(14.180938,2.7596874)
\psline[linewidth=0.04cm,arrowsize=0.05291667cm 2.0,arrowlength=1.4,arrowinset=0.4]{->}(14.180938,1.8596874)(14.180938,1.4596875)
\psline[linewidth=0.04cm,arrowsize=0.05291667cm 2.0,arrowlength=1.4,arrowinset=0.4]{->}(14.180938,-0.8403125)(14.180938,-1.2403125)
\psline[linewidth=0.04cm](14.380938,2.5596876)(15.280937,2.5596876)
\psline[linewidth=0.04cm](14.380938,-1.4403125)(15.280937,-1.4403125)
\psline[linewidth=0.04cm](14.380938,1.2596875)(15.280937,1.2596875)
\psframe[linewidth=0.04,dimen=outer](3.3809376,3.0596876)(1.9809375,-1.9403125)
\usefont{T1}{ptm}{m}{n}
\rput{-90.0}(1.9128125,3.4043748){\rput(2.6567187,0.7696875){User Scheduling }}
\usefont{T1}{ptm}{m}{n}
\rput(3.6223438,2.9696875){$\mathbf u_1$}
\psline[linewidth=0.04cm](3.9809375,1.3596874)(3.7809374,1.1596875)
\usefont{T1}{ptm}{m}{n}
\rput(3.8823438,0.9696875){$n_r$}
\usefont{T1}{ptm}{m}{n}
\rput(3.6223438,1.6696875){$\mathbf u_2$}
\usefont{T1}{ptm}{m}{n}
\rput(3.6623437,-1.1303124){$\mathbf u_K$}
\usefont{T1}{ptm}{m}{n}
\rput(5.0023437,2.5696876){$\mathbf W_1$}
\usefont{T1}{ptm}{m}{n}
\rput(5.0023437,1.2696875){$\mathbf W_2$}
\usefont{T1}{ptm}{m}{n}
\rput(5.0423436,-1.4303125){$\mathbf W_K$}
\usefont{T1}{ptm}{m}{n}
\rput(6.3123436,2.7696874){$\mathbf x_1$}
\psline[linewidth=0.04cm](6.4809375,2.0596876)(6.2809377,1.8596874)
\usefont{T1}{ptm}{m}{n}
\rput(6.7123437,1.9696875){$N_t$}
\usefont{T1}{ptm}{m}{n}
\rput(6.7123437,-0.4303125){$N_t$}
\psline[linewidth=0.04cm](6.4809375,-0.3403125)(6.2809377,-0.5403125)
\usefont{T1}{ptm}{m}{n}
\rput(6.3523436,-1.6303124){$\mathbf x_K$}
\usefont{T1}{ptm}{m}{n}
\rput(6.8323436,0.8696875){$\mathbf x$}
\psline[linewidth=0.04cm](6.9809375,0.6596875)(6.7809377,0.4596875)
\usefont{T1}{ptm}{m}{n}
\rput(6.8123436,0.2696875){$N_t$}
\usefont{T1}{ptm}{m}{n}
\rput(8.632343,2.5696876){$\text{BS}_1$}
\usefont{T1}{ptm}{m}{n}
\rput(8.632343,1.2696875){$\text{BS}_2$}
\usefont{T1}{ptm}{m}{n}
\rput(8.662344,-1.4303125){$\text{BS}_B$}
\psline[linewidth=0.04cm](10.580937,0.6596875)(10.380938,0.4596875)
\usefont{T1}{ptm}{m}{n}
\rput(10.512343,0.8696875){$N_t$}
\usefont{T1}{ptm}{m}{n}
\rput(15.9323435,2.5696876){$\mathbf F_1$}
\usefont{T1}{ptm}{m}{n}
\rput(15.9323435,1.2696875){$\mathbf F_2$}
\usefont{T1}{ptm}{m}{n}
\rput(15.972343,-1.4303125){$\mathbf F_K$}
\psline[linewidth=0.04cm](10.080937,2.1596875)(9.880938,1.9596875)
\usefont{T1}{ptm}{m}{n}
\rput(10.372344,2.0696876){$n_t$}
\usefont{T1}{ptm}{m}{n}
\rput(10.272344,-0.4303125){$n_t$}
\psline[linewidth=0.04cm](10.080937,-0.3403125)(9.880938,-0.5403125)
\usefont{T1}{ptm}{m}{n}
\rput(12.462344,2.5696876){$\mathbf H_1$}
\usefont{T1}{ptm}{m}{n}
\rput(12.502344,-1.4303125){$\mathbf H_K$}
\usefont{T1}{ptm}{m}{n}
\rput(12.462344,1.2696875){$\mathbf H_2$}
\usefont{T1}{ptm}{m}{n}
\rput(14.122344,3.4696875){$\mathbf n_1$}
\usefont{T1}{ptm}{m}{n}
\rput(14.122344,2.0696876){$\mathbf n_2$}
\usefont{T1}{ptm}{m}{n}
\rput(14.162344,-0.6303125){$\mathbf n_K$}
\usefont{T1}{ptm}{m}{n}
\rput(14.722343,2.3696876){$\mathbf y_1$}
\usefont{T1}{ptm}{m}{n}
\rput(14.722343,0.9696875){$\mathbf y_2$}
\usefont{T1}{ptm}{m}{n}
\rput(14.662344,-1.7303125){$\mathbf y_K$}
\psframe[linewidth=0.04,linestyle=dashed,dash=0.16cm 0.16cm,dimen=outer](16.880938,3.5596876)(15.080937,-2.2403126)
\usefont{T1}{ptm}{m}{n}
\rput(15.939688,-2.6303124){inter-cluster interference}
\psframe[linewidth=0.04,linestyle=dashed,dash=0.16cm 0.16cm,dimen=outer](6.0809374,3.4596875)(4.0809374,-2.3403125)
\usefont{T1}{ptm}{m}{n}
\rput(5.0823436,-2.8303125){$\left[\sum_{k=1}^K\mathbf W_k\mathbf W_k^\textsf{H}\right]_{ii} \leq p_i$}
\usefont{T1}{ptm}{m}{n}
\rput(15.874063,-3.0303125){cancelation}
\usefont{T1}{ptm}{m}{n}
\rput(5.032344,-3.4303124){$i=1,\ldots,N_t$}
\psframe[linewidth=0.04,linestyle=dashed,dash=0.16cm 0.16cm,dimen=outer](9.680938,3.4596875)(7.6809373,-2.3403125)
\usefont{T1}{ptm}{m}{n}
\rput(8.603594,-2.7303126){Coordination}
\psline[linewidth=0.04cm](14.780937,2.6596875)(14.580937,2.4596875)
\usefont{T1}{ptm}{m}{n}
\rput(14.6823435,2.8696876){$n_r$}
\psline[linewidth=0.04cm](14.780937,1.3596874)(14.580937,1.1596875)
\usefont{T1}{ptm}{m}{n}
\rput(14.6823435,1.5696875){$n_r$}
\psline[linewidth=0.04cm](14.780937,-1.3403125)(14.580937,-1.5403125)
\usefont{T1}{ptm}{m}{n}
\rput(14.6823435,-1.1303124){$n_r$}
\psline[linewidth=0.04cm,arrowsize=0.05291667cm 2.0,arrowlength=1.4,arrowinset=0.4]{->}(16.680937,-1.4403125)(17.780937,-1.4403125)
\psline[linewidth=0.04cm](17.280937,-1.3403125)(17.080938,-1.5403125)
\usefont{T1}{ptm}{m}{n}
\rput(18.132343,-1.4303125){$\mathbf r_K$}
\usefont{T1}{ptm}{m}{n}
\rput(17.182344,-1.1303124){$n_r$}
\usefont{T1}{ptm}{m}{n}
\rput(18.092344,1.2696875){$\mathbf r_2$}
\usefont{T1}{ptm}{m}{n}
\rput(17.182344,1.5696875){$n_r$}
\psline[linewidth=0.04cm](17.280937,1.3596874)(17.080938,1.1596875)
\psline[linewidth=0.04cm,arrowsize=0.05291667cm 2.0,arrowlength=1.4,arrowinset=0.4]{->}(16.680937,1.2596875)(17.780937,1.2596875)
\psline[linewidth=0.04cm,arrowsize=0.05291667cm 2.0,arrowlength=1.4,arrowinset=0.4]{->}(16.680937,2.5596876)(17.780937,2.5596876)
\usefont{T1}{ptm}{m}{n}
\rput(18.132343,2.5696876){$\mathbf r_K$}
\usefont{T1}{ptm}{m}{n}
\rput(17.182344,2.7696874){$n_r$}
\psline[linewidth=0.04cm](17.280937,2.6596875)(17.080938,2.4596875)
\psdots[dotsize=0.12](8.680938,-0.4403125)
\psdots[dotsize=0.12](8.680938,-0.1403125)
\psdots[dotsize=0.12](8.680938,0.1596875)
\psdots[dotsize=0.12](12.480938,-0.4403125)
\psdots[dotsize=0.12](12.480938,-0.1403125)
\psdots[dotsize=0.12](12.480938,0.1596875)
\psdots[dotsize=0.12](16.080938,-0.4403125)
\psdots[dotsize=0.12](16.080938,-0.1403125)
\psdots[dotsize=0.12](16.080938,0.1596875)
\end{pspicture}
}

\caption{Block diagram of network MIMO coordination transmission strategy.}
\label{fig:Schematic}
\end{figure*}

A block diagram of transmission strategy for network MIMO coordination is shown in Fig~\ref{fig:Schematic}.
The transmitted symbol to user $k$ is an $n_r$-dimensional
vector $\mathbf u_k$, which is multiplied by a $N_t \times n_r$
precoding matrix $\mathbf W_k$ and passed on to the base station's
antenna array. Since all base station antennas are coordinated, the
complex antenna output vector $\mathbf x$ is composed of signals for
all $K$ users. Therefore, $\mathbf x$ can be written as follows
\begin{equation}\label{TxRxequation}
\mathbf x=\sum_{k=1}^K\mathbf W_k\mathbf u_k
\end{equation} where $\mathbb E[\mathbf u_k\mathbf u_k^\textsf{H}]=\mathbf I_{n_r}$. The received signal $\mathbf y_k$ at user $k$ can
be represented as
\begin{equation}\label{ReceivedSignal3}
\mathbf y_k=\mathbf H_k\mathbf W_k\mathbf u_k+\sum_{j
\neq k}\mathbf H_k \mathbf W_j \mathbf u_j+ \mathbf z_k,
\end{equation}
where $\mathbf z_k \thicksim \mathcal {CN}(0,\mathbf I_{n_r})$ denotes the normalized AWGN vector at user $k$. The random characteristics of channel matrix entries of $\mathbf H_{k}$ are discussed in Section \ref{Sec:SimulationSetup}. They encompass three factors: path loss, Rayleigh fading, and log-normal shadowing. Random structure of the channel coefficients ensures $\text{rank}(\mathbf H_{k})=\min(n_r,N_t)=n_r$ for user $k$ with probability one. 
Per-antenna power constraints (\ref{PAPC}) impose a power constraint
\begin{equation}
\begin{split}
[\mathbf S_x]_{i,i}=&\mathbb E[\mathbf x \mathbf x^\textsf{H}]_{i,i}\\=&\left[\sum_{k=1}^K\mathbf W_k\mathbf W_k^\textsf{H}\right]_{i,i}
\leq p_i, \quad i=1,\ldots, N_t
\end{split}
\end{equation}
on each transmit antenna. The sum power constraint also can be expressed as
\begin{equation}
\tr\left\{\mathbf S_x\right\}=\sum_{k=1}^K \tr\left\{\mathbf W_k
\mathbf W_k^\textsf{H}\right\} \leq P.
\end{equation}
\par Due to the structure of multiuser zero-forcing scheme, the number of users that can be served simultaneously in each time slot is limited. Hence, user selection algorithm is necessary. We consider two main criteria for the user selection scheme: maximum sum rate (MSR) and proportional fairness. We employ the greedy user selection algorithm discussed in \cite{Dimic05,Shen08}. The proportionally-fair user selection algorithm is based on greedy weighted user selection algorithm with an update of the weights discussed in \cite{Viswanathan03,Sigdel09,Sigdel08}.

\section{Multi-cell Multiuser Block Diagonalization}\label{Sec:Multi-CellBD}
\par To remove the intra-cluster interference, a practical linear zero-forcing can be employed. Applying the multiuser zero-forcing to the multiple-antenna users requires block diagonalization rather than channel inversion \cite{Spencer04}. Assuming the transmission strategy in Section \ref{Sec:TransmissionStrategy}, each user's data $\mathbf u_k$ is precoded with the matrix $\mathbf W_k$, such that
\begin{equation}\label{orthogonaltx}
\mathbf H_k \mathbf W_j=0 \quad \text{for all} \quad k \neq j \quad
\text{and} \quad 1 \leq k,j \leq K.
\end{equation}
Hence the received signal for user $k$ can be simplified
to
\begin{equation}\label{receivedSignal2}
\mathbf y_k=\mathbf H_k\mathbf W_k\mathbf u_k+\mathbf n_k.
\end{equation}
Let $\widetilde{\mathbf H}_k=[\mathbf H_1^\textsf{T} \cdots \mathbf H_{k-1}^\textsf{T}
\mathbf H_{k+1}^\textsf{T} \cdots \mathbf H_K^\textsf{T}]^\textsf{T}$. Zero-interference constraint in
(\ref{orthogonaltx}) forces $\mathbf W_k$ to lie in the null space
of $\widetilde{\mathbf H}_k$ which requires a dimension condition
$Bn_t \geq Kn_r$ to be satisfied. This directly comes from the definition of null space in linear algebra \cite{Horn85}.
Hence, the maximum number of users that can be served in a time slot is $K=\lfloor\frac{N_t}{n_r}\rfloor$. We focus on the $K$ users which are selected through a scheduling algorithm and assigned to one subband. The remaining unserved users are referred to other subbands or will be scheduled in other time slots. Recall that the vectors in (\ref{TxRxequation}) are normalized with respect to the standard deviation of the additive noise component, $\sigma$, resulting in $\mathbf n_k$ having components with unit variance. Assume that $\widetilde{\mathbf H}_k$ is a full rank matrix
\mbox{$\text{rank}(\widetilde{\mathbf H}_k)=(K-1)n_r$}, which holds with probability one due to the randomness of entries of channel matrices. We perform singular value decomposition (SVD)
\begin{equation}\label{SVD}
\widetilde{\mathbf H}_k=\mathbf U_k\mathbf
\Lambda_k \left[\mathbf \Upsilon_k \mathbf
V_k\right]^\textsf{T}
\end{equation}
where $\mathbf \Upsilon_k$ holds the first
$(K-1)n_r$ right singular vectors corresponding to non-zero singular values, and $\mathbf
V_k \in \mathbb C^{N_t \times m_r}$ contains the last $m_r=N_t-(K-1)n_r$ right
singular vectors corresponding to zero singular values of $\widetilde{\mathbf H}_k$. If number of scheduled users is $K=\frac{N_t}{n_r}$ then $m_r=n_r$, otherwise $m_r > n_r$ when $K < \frac{N_t}{n_r}$. The orthonormality of $\mathbf V_k$ means that $\mathbf V_k^\textsf{H}\mathbf V_k=\mathbf I_{m_r}$. The columns of
$\mathbf V_k$ form a basis set in the null space
of $\widetilde{\mathbf H}_k$, and hence $\mathbf W_k$ can be any
linear combination of the columns of $\mathbf V_k$, i.e.
\begin{equation}\label{PrecoderDesign}
\mathbf W_k=\mathbf V_k \mathbf \Psi_k, \quad k=1,\ldots, K
\end{equation}
where $\mathbf \Psi_k \in \mathbb C^{m_r \times n_r}$ can be any arbitrary matrix subject to the per-antenna power constraints \cite{Kaviani08}. Conventional BD scheme proposed in \cite{Spencer04} assumes only linear combinations of a diagonal form to simplify it to a power allocation algorithm through water-filling. The conventional BD is optimal only when sum power constraint is applied \cite{Kaviani09}, and it is not optimal under per-antenna power constraints \cite{karakayali07,Wiesel08,Huh09}.
\subsection{Conventional BD}
In conventional BD \cite{Spencer04}, the sum power constraint is applied to the throughput maximization problem and further relaxed to a simple water-filling power allocation algorithm. In this scheme, the linear combination introduced in (\ref{PrecoderDesign}) is confined to have a form given by
\begin{equation}
\mathbf \Psi_k= \widetilde{\mathbf V}_k\mathbf \Theta_k^{\frac{1}{2}}, \quad k=1,\ldots,K
\end{equation}
where $ \widetilde{\mathbf V}_k \in \mathbb C^{m_r \times n_r}$ are the right singular vectors of $\mathbf H_k \mathbf V_k$ corresponding to its non-zero singular values. Hence, the aggregate precoding matrix of the conventional scheme, $\mathbf W_{\text{BD}}$, is defined as
\begin{equation}
\mathbf W_{\text{BD}}=\left[\mathbf V_1\widetilde{\mathbf V}_1  \quad \mathbf V_2\widetilde{\mathbf V}_2 \quad \cdots \quad \mathbf V_K\widetilde{\mathbf V}_K\right]\mathbf \Theta^{\frac{1}{2}}
\end{equation}
where $\mathbf \Theta=\bdiag\left[\mathbf \Theta_1 , \cdots ,\mathbf \Theta_K\right]$ is a diagonal matrix whose elements scale the power transmitted into each of the columns of $\mathbf W_{\text{BD}}$. The sum power constraint implies that
\begin{equation}
\sum_{k=1}^K \tr\left\{ \mathbf V_k \widetilde{\mathbf V}_k \mathbf \Theta_k \widetilde{\mathbf V}_k^\textsf{H} \mathbf V_k^\textsf{H}\right\}=\sum_{k=1}^K\tr\left\{\mathbf \Theta_k\right\}
\end{equation}
This relaxes the problem to optimization over the diagonal terms of $\mathbf \Theta_k$ and consequently is interpreted as a power allocation problem and can be solved through well-known water-filling algorithm over the diagonal terms of $\mathbf \Theta$. However, this form of BD cannot be extended as an optimal precoder to the case of per-antenna power constraints because
\begin{equation}\label{PAPCneqTSPC}
\left[\mathbf W_{\text{BD}}\mathbf W_{\text{BD}}^\textsf{H}\right]_{i,i}=\left[ \mathbf V_{\text{BD}}\mathbf \Theta \mathbf V_{\text{BD}}^\textsf{H}\right]_{i,i} \neq \left[\mathbf \Theta\right]_{i,i},
\end{equation}
where $\mathbf V_{\text{BD}}=\left[\mathbf V_1\widetilde{\mathbf V}_1  \quad \mathbf V_2\widetilde{\mathbf V}_2 \quad \cdots \quad \mathbf V_K\widetilde{\mathbf V}_K\right]$. Note that $i$th diagonal term of the left side of (\ref{PAPCneqTSPC}) is a linear combination of all entries of matrix $\mathbf \Theta$ and not only the diagonal terms. The selection of $\mathbf \Theta$ as a diagonal matrix reduces the search domain size of optimization and hence does not lead to optimal solution. Furthermore, $\widetilde{\mathbf V}_k$ impacts the diagonal terms of $\mathbf W_{\text{BD}}\mathbf W_{\text{BD}}^\textsf{H}$ (i.e. transmission covariance matrix) and therefore insertion of $\widetilde{\mathbf V}_k$ not necessarily reduces the required power allocated to each antenna. In addition it adds $K$ SVD operations to the precoding computation procedure (one for each served users) to find $\widetilde{\mathbf V}_k$. Additionally, the per-antenna power constraints do not allow the optimization to lead to simple water-filling algorithm. Previous work on BD with per-antenna (similarly with per-base-station) power constraints for a case of multiple-receive antennas employs this conventional BD and optimizes diagonal terms of $\mathbf \Theta$ \cite{Zhang09,karakayali07,Hadisusanto08}. Hence, it is not optimal. The optimal form of BD proposed in this paper includes the optimization over the entire null space of other users' channel matrices resulting in optimal precoders under per-antenna power constraints, easily extendable to per-base station power constraints.
\begin{figure}[!t]
\centering
 \includegraphics[scale=.4]{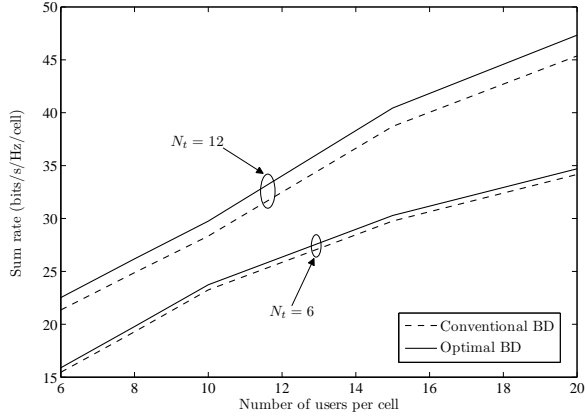}
\caption{Comparison of sum rates for conventional BD vs. the proposed optimal BD for $B=1$, $N_t=6, 12$, $n_r=2$ using maximum sum rate scheduling.}
\label{fig:OptimalvsCoventional}
\end{figure}
 \par The numerical results in Fig.~\ref{fig:OptimalvsCoventional} compare maximized sum rate of a MIMO BC system with conventional BD \cite{Spencer04} and the optimal scheme proposed later in this paper. There are 12 transmit antennas at the base station and 2 receive antennas at each mobile user. $B=1$ is considered to specifically show the difference between the two BD schemes. Note that the conventional BD has a domain of $\mathbb R_{+}^{N_t}$ while the optimal BD searches over all possible $K$ symmetric matrices of $\mathbf \Phi_k$ and therefore has a larger domain of $\mathbb C_{++}^{Kn_r(n_r-1)/2}$ and grows when number of users per cell increases. As a consequence, the difference between these two schemes increases with the number of users per cell. Details of the simulation setup are given in Section \ref{Sec:SimulationSetup}. In the following section the optimal BD scheme is introduced and discussed in detail, and the algorithm to find the precoders is presented.

\subsection{Optimal Multi-Cell BD}\label{Sec:OptimalBD}
 The focus of this section is on the design of optimal multi-cell BD precoder matrices $\mathbf W_k$ to maximize the throughput while enforcing per-antenna power constraints. In this scheme, we search over the entire null space of other users channel matrices ($\widetilde{\mathbf H}_k$), i.e. $\mathbf \Psi_k$ can be any arbitrary matrix of $\mathbb C^{m_r \times n_r}$ satisfying the per-antenna power constraints.

\par Following the design of precoders according to (\ref{PrecoderDesign}), the received signal for user $k$ can be expressed as
\begin{equation}
\mathbf y_k=\mathbf H_k\mathbf V_k\mathbf \Psi_k\mathbf u_k+\mathbf z_k.
\end{equation}
Denote $\mathbf \Phi_k=\mathbf \Psi_k \mathbf \Psi_k^\textsf{H} \in \mathbb C^{m_r \times m_r}$, \mbox{$k=1, \ldots, K$}, which are positive semi-definite matrices. The rate of $k$th user is given by
\begin{equation}
R_k=\log \left|\mathbf I+\mathbf H_k\mathbf V_k\mathbf \Phi_k \mathbf V^\textsf{H}_k\mathbf H_k^\textsf{H}\right|.
\end{equation}
Therefore, sum rate maximization problem can be expressed as
\begin{equation}\label{optimization1}
\begin{array}{rl}
\text{maximize}& \sum_{k=1}^K\log\left|\mathbf I+\mathbf H_k\mathbf V_k\mathbf \Phi_k \mathbf V^{\textsf{H}}_k\mathbf H_k^\textsf{H}\right|\\
 \text{subject to} & \left[\sum_{k=1}^K\mathbf V_k\mathbf \Phi_k \mathbf V^\textsf{H}_k\right]_{i,i} \leq p_i,\quad  i=1,\ldots,N_t\\
&  \mathbf \Phi_k \succeq 0, \quad k=1,\ldots,K,
\end{array}
\end{equation}
where the maximization is over all positive semi-definite matrices $\mathbf \Phi_1, \ldots, \mathbf \Phi_K$ with a rank constraint of $\rank(\mathbf \Phi_k) \leq n_r$. Notice that the objective function in (\ref{optimization1}) is concave \cite[p. 466]{Horn85} and the constraints are also affine functions \cite{Boyd04}. Thus, the problem is categorized as a convex optimization problem. We propose a gradient descent algorithm to find the optimal BD precoders. We define $\mathbf G_k=\mathbf H_k\mathbf V_k$ and correspondingly its right pseudo-inverse matrix as $\mathbf G_k^\dag=\mathbf G_k^\mathsf H \left(\mathbf G_k \mathbf G_k^\mathsf H\right)^{-1}$. Let $\mathbf Q_k=\mathbf V_k\mathbf G_k^{-1}$ which is an $N_t \times n_r$ matrix and we perform the SVD $\mathbf Q_k^\textsf{H} \mathbf \Lambda \mathbf Q_k=\mathbf U_{k}\mathbf \Sigma_k \mathbf U_k^\textsf{H}$. We introduce the positive semi-definite matrices $\mathbf \Omega_k$ defined as
\begin{equation}\label{Omega_k}
\mathbf \Omega_k=\mathbf U_k\left[\mathbf \Sigma_k-\mathbf I\right]_+\mathbf U_k^\textsf{H},
\end{equation}
where the operator $\left[\mathbf D\right]_+=\diag\left[\max(0,d_{1}),\ldots,\max(0,d_n)\right]$ on a diagonal matrix $\mathbf D=\diag\left[d_1,\ldots,d_n\right]$.
\par
\begin{Theorem}
The optimal BD precoders can be obtained through solving the dual problem
\begin{equation}\label{DualProblem}
\begin{array}{cl}
\text{minimize}& g(\mathbf \Lambda)\\
 \text{subject to} & \mathbf \Lambda \succeq 0, \quad \mathbf \Lambda \text{ diagonal}
\end{array}
\end{equation}
where
\begin{equation}\label{gFunction}
\begin{split}
g\left(\mathbf \Lambda\right)=& -\sum_{k=1}^K \log\left|\mathbf Q_k^\textsf{H}\mathbf \Lambda \mathbf Q_k -\mathbf \Omega_k\right|-Kn_r\\
& +\tr\left\{\sum_{k=1}^K\left(\mathbf Q_k^\textsf{H}\mathbf \Lambda \mathbf Q_k -\mathbf \Omega_k\right)\right\}+\tr\left\{\mathbf \Lambda\mathbf P\right\}.
\end{split}
\end{equation}
with a gradient descent direction given as
\begin{equation}\label{GradientDescentSearchDirection}
\begin{split}
\Delta \mathbf \Lambda=&\sum_{k=1}^K \diag\left[\mathbf Q_k \left(\mathbf Q_k^\textsf{H} \mathbf \Lambda \mathbf Q_k-\mathbf \Omega_k\right)^{-1}\mathbf Q_k^\textsf{H}\right]\\&-\mathbf P-\sum_{k=1}^K \diag\left[\mathbf Q_k \mathbf Q_k^\textsf{H}\right].
\end{split}
\end{equation}
The optimal BD precoders for the optimal value of $\mathbf \Lambda^\star$ is given as
\begin{equation}\label{eqn:OptimalBDPrecoder}
\mathbf W_k=\mathbf V_k\left[\mathbf G_k^{\dag}\left(\left(\mathbf Q_k^\textsf{H}\mathbf \Lambda^\star \mathbf Q_k-\mathbf \Omega_k\right)^{-1}-\mathbf I\right)\left(\mathbf G_k^{\dag}\right)^\mathsf H\right]^{\frac{1}{2}}.
\end{equation}
\end{Theorem}

\begin{IEEEproof}
The proof is given in the Appendix.

\end{IEEEproof}

\par The KKT conditions for the dual problem are given as
\begin{equation}\label{StoppingCriterion}
\begin{split}
\mathbf \Lambda \succeq & 0,\\
\nabla _{ \mathbf \Lambda}g \succeq & 0,\\
\lambda_i\left[\nabla_{\mathbf \Lambda}g\right]_{i,i}= & 0, \quad i=1,\ldots, N_t
\end{split}
\end{equation}
with the last condition being the complementarity \cite[p. 142]{Boyd04}.
Thus, the stopping criterion for the gradient descent method can be established using small values of $\epsilon \geq 0$ replacing zero values.
\par More interestingly, the sum rate maximization in (\ref{optimization1}) through dual problem in (\ref{DualProblem}) facilitates the extension to any linear power constraints on the transmit antennas. The dual problem has $N_t$ variables $\lambda_i$, $i=1,\ldots,N_t$, one for each transmit antenna power constraint. More general power constraints than those given in (\ref{optimization1})can be defined as \cite{Huh09}
\begin{equation}\label{LinearConstraints}
\tr\left\{\sum_{k=1}^K \mathbf V_k \mathbf \Phi_k \mathbf V_k^\textsf{H} \mathbf T_l\right\}\leq p_l, \quad l=1,\ldots,L
\end{equation}
where $\mathbf T_l$ are positive semidefinite symmetric matrices and $p_l$ are non-negative values corresponding to each of $L$ linear constraints. The special case of this structure of power constraints has been discussed frequently in the literature: for $L=1$, $p_1=P$ and $\mathbf T_1=\mathbf I$ the conventional sum power constraint results \cite{Spencer04}; when $L=N_t$ and $\mathbf T_l$ is a matrix with its $l$th diagonal term equal to one and all other elements zero, we get per antenna power constraints studied in this section. Another scenario is per-base station power constraint, which is derived with $L=B$, $p_l=P_l$ ($l$th per-base power limit) and $\mathbf T_l$ all zero except equal to one on $n_t$ terms of its diagonal each corresponding to one of the $l$th base station's transmit antennas. When the sum power constraint is applied only one dual variable is needed in dual optimization problem (\ref{DualProblem}) (i.e. $\mathbf \Lambda=\lambda \mathbf I_{N_t}$), where $\lambda$ determines the water level in the water-filling algorithm \cite{Spencer04}. For per-base station power constraints, the optimization dual variable can be defined as $\mathbf \Lambda= \mathbf \Lambda_{bs} \otimes \mathbf I_{n_t}$,
where $\mathbf \Lambda_{bs}=\diag\left[\lambda_1,\ldots,\lambda_{B}\right]$ consists of $B$ dual variables (one for each base station) and the operator $\otimes$ is the Kronecker product \cite{Horn85}. The details of the optimization steps in the per-base station power constraints scenario are discussed in Section \ref{PerBaseStationPowerConstraintsScenario} and the study of general linear constraints is left for further work.

\subsection{Per-Base-Station Power Constraints}\label{PerBaseStationPowerConstraintsScenario}
In this Section, the extension of the ZF beamforming optimization to the system with per-base station power constraint is considered. The optimization problem in (\ref{optimization1}) can be rewritten considering the per-base-station power constraints as
\begin{equation}\label{optimization6}
\begin{array}{rl}
\text{maximize}& \sum_{k=1}^K\log\left|\mathbf I+\mathbf H_k\mathbf V_k\mathbf \Phi_k \mathbf V^\textsf{H}_k\mathbf H_k^\textsf{H}\right|\\
 \text{subject to} & \tr\left\{\mathbf \Delta_b\left(\sum_{k=1}^K\mathbf V_k\mathbf \Phi_k \mathbf V^\textsf{H}_k\right)\right\} \leq P_b,\\
 & \quad  b=1,\ldots,B\\
&  \mathbf \Phi_k \succeq 0, \quad k=1,\ldots,K.
\end{array}
\end{equation}
where $P_1, \ldots, P_B$ are the per-base station maximum powers and $\mathbf \Delta_b$ is a diagonal matrix with its entries equal to one for the corresponding antennas within the base-station $b$ and the rest equal to zero. For the simplicity, $b$th $n_t$-entries of the diagonal of $\mathbf \Delta_b$ are only equal to one. Following similar steps as (\ref{optimization2}), the Lagrange dual function is obtained as
\begin{equation}
\begin{split}
\mathcal L(\left\{\mathbf S\right\},\mathbf \lambda)=&\sum_{k=1}^K\log\left|\mathbf I+\mathbf S_k\right|\\
&-\!\sum_{b=1}^B\tr\left\{\!\lambda_b\mathbf \Delta_b\left(\!\sum_{k=1}^K\mathbf Q_k\mathbf S_k \mathbf Q^\textsf{H}_k\!-\!\mathbf P_{bs} \otimes \mathbf I_{n_t}\!\right)\!\right\}\\
&+\sum_{k=1}^K \tr\left\{\mathbf \Omega_k \mathbf S_k\right\}
\end{split}
\end{equation}
where $\mathbf P_{bs}=\diag\left[P_1,\ldots,P_{B}\right]$ and $\otimes$ is the Kronecker product \cite{Horn85}.
The KKT conditions yield that
\begin{equation}
\mathbf S_k\!=\!\left[\mathbf Q_k^\textsf{H}\!\left(\mathbf \Lambda_{bs} \otimes \mathbf I_{n_t} \right) \!\mathbf Q_k\!-\mathbf \Omega_k\right]^{-1}\!-\mathbf I, \quad k=1,\ldots, K
\end{equation}
where $\mathbf \Lambda_{bs}=\diag\left[\lambda_1,\ldots,\lambda_B\right]$ and $\mathbf \Omega_k$ can be defined in a similar way as (\ref{Omega_k}). The dual problem can be expressed similarly as (\ref{DualProblem}). Following the steps in Section \ref{Sec:OptimalBD}, the gradient descent search direction is given by
\begin{equation}
\begin{split}
&\nabla_{\mathbf \Lambda}g=\\&\!-\sum_{k=1}^K \diag_{b=1,\ldots,B}\!\left[\tr_{b}\left\{\mathbf Q_k \left[\mathbf Q_k^\textsf{H} \left(\mathbf \Lambda_{bs} \otimes \mathbf I_{n_t} \right) \mathbf Q_k-\mathbf \Omega_k\right]^{-1}\mathbf Q_k^\textsf{H}\right\}\right]\\&+\mathbf P_{bs}+\sum_{k=1}^K \diag_{b=1,\ldots,B}\left[\tr_{b}\left\{\mathbf Q_k \mathbf Q_k^\textsf{H}\right\}\right]
\end{split}
\end{equation}
where $\tr_b$ is a partial matrix trace over $b$th $n_t$-entries of the diagonal terms of a matrix. $\diag_{b=1,\ldots,B}\left[\cdot \right]$ gives a diagonal matrix with $B$ elements computed for each $b=1,\ldots,B$.

\subsection{Single Antenna Receivers}
Although this paper studies a network MIMO system with multiple receive antenna users, the results can be applied to a system with single receive antenna users. In this case each user's transmission must be orthogonal to a vector (rather than a matrix), which is the basis vector for other users' transmissions. The optimization is over all real vectors with positive elements ($\mathbb R_+^{N_t}$) satisfying the power constraints. This approach facilitates the optimization presented in \cite{Wiesel08} and \cite{Huh09} using the generalized inverses and multi-step optimizations.

\section{Simulation Setup}\label{Sec:SimulationSetup}
The propagation model between each base station's transmit antenna and mobile user's receive antenna includes three factors: a path loss component proportional to $d_{kb}^{-\beta}$ (where $d_{kb}$ denotes distance from base station $b$ to mobile user $k$ and
$\beta=3.8$ is the path loss exponent), and two random components representing lognormal shadow fading and
Rayleigh fading. The channel gain between transmit antenna $t$ of the base station $b$ and receive antenna $r$ of the $k$th user is given by
\begin{equation}
\left[\mathbf H_{k,b}\right]_{(r,t)}=\alpha_{k,b}^{(r,t)}\sqrt{\rho_{k,b}\left(\frac{d_{kb}}{d_0}\right)^{-\beta} \mathit{\Gamma}}
\end{equation}
where $\left[\mathbf H_{k,b}\right]_{(r,t)}$ is the $(r,t)$ element of the channel matrix $\mathbf H_{k,b} \in \mathbb C^{n_r \times n_t}$ from the base station $b$ to the mobile user $k$, $\alpha_{k,b}^{(r,t)} \sim \mathcal{CN}\left(0,1\right)$ represents independent Rayleigh fading, $d_0=1 \text{ km}$ is the cell radius, and $\rho_{k,b}=10^{\rho_{k,b}^\text{(dBm)}/10}$ is the lognormal shadow fading variable between $b$th base station and $k$th user, where $\rho_{k,b}^{(\text{dBm})} \sim \mathcal{CN}\left(0,\sigma_{\rho}\right)$ and  $\sigma_{\rho}=8 \text{ dB}$ is its standard deviation. A reference SNR, $\mathit{\Gamma}=20 \text{ dB}$ is a typical value of the interference-free SNR at the cell boundary (as in \cite{Karakayali06d} and \cite{Huang09}).
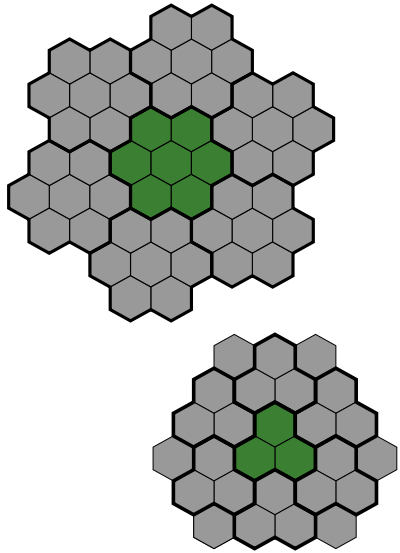
\begin{figure}
\centering
\scalebox{.3} {
\begin{pspicture}(0,-13)(17.22,13)
\definecolor{color360b}{rgb}{0.0,0.6,0.2}
\definecolor{color422b}{rgb}{0.6,0.6,0.6}
\pspolygon[linewidth=0.05,fillstyle=solid,fillcolor=color422b](3.6,9.05)(2.7,8.55)(2.7,7.55)(3.6,7.05)(4.5,7.55)(4.5,8.55)(3.6,9.05)
\pspolygon[linewidth=0.05,fillstyle=solid,fillcolor=color422b](4.5,10.55)(3.6,10.05)(3.6,9.05)(4.5,8.55)(5.4,9.05)(5.4,10.05)(4.5,10.55)
\pspolygon[linewidth=0.05,fillstyle=solid,fillcolor=color422b](4.5,7.55)(3.6,7.05)(3.6,6.05)(4.5,5.55)(5.4,6.05)(5.4,7.05)(4.5,7.55)
\pspolygon[linewidth=0.05,fillstyle=solid,fillcolor=color422b](5.4,9.05)(4.5,8.55)(4.5,7.55)(5.4,7.05)(6.3,7.55)(6.3,8.55)(5.4,9.05)
\pspolygon[linewidth=0.05,fillstyle=solid,fillcolor=color422b](2.7,10.55)(1.8,10.05)(1.8,9.05)(2.7,8.55)(3.6,9.05)(3.6,10.05)(2.7,10.55)
\pspolygon[linewidth=0.05,fillstyle=solid,fillcolor=color422b](1.8,9.05)(0.9,8.55)(0.9,7.55)(1.8,7.05)(2.7,7.55)(2.7,8.55)(1.8,9.05)
\pspolygon[linewidth=0.05,fillstyle=solid,fillcolor=color422b](2.7,7.55)(1.8,7.05)(1.8,6.05)(2.7,5.55)(3.6,6.05)(3.6,7.05)(2.7,7.55)
\pspolygon[linewidth=0.05,fillstyle=solid,fillcolor=OliveGreen](7.2,6.05)(6.3,5.55)(6.3,4.55)(7.2,4.05)(8.1,4.55)(8.1,5.55)(7.2,6.05)
\pspolygon[linewidth=0.05,fillstyle=solid,fillcolor=OliveGreen](8.1,7.55)(7.2,7.05)(7.2,6.05)(8.1,5.55)(9.0,6.05)(9.0,7.05)(8.1,7.55)
\pspolygon[linewidth=0.05,fillstyle=solid,fillcolor=OliveGreen](8.1,4.55)(7.2,4.05)(7.2,3.05)(8.1,2.55)(9.0,3.05)(9.0,4.05)(8.1,4.55)
\pspolygon[linewidth=0.05,fillstyle=solid,fillcolor=OliveGreen](9.0,6.05)(8.1,5.55)(8.1,4.55)(9.0,4.05)(9.9,4.55)(9.9,5.55)(9.0,6.05)
\pspolygon[linewidth=0.05,fillstyle=solid,fillcolor=OliveGreen](6.3,7.55)(5.4,7.05)(5.4,6.05)(6.3,5.55)(7.2,6.05)(7.2,7.05)(6.3,7.55)
\pspolygon[linewidth=0.05,fillstyle=solid,fillcolor=OliveGreen](5.4,6.05)(4.5,5.55)(4.5,4.55)(5.4,4.05)(6.3,4.55)(6.3,5.55)(5.4,6.05)
\pspolygon[linewidth=0.05,fillstyle=solid,fillcolor=OliveGreen](6.3,4.55)(5.4,4.05)(5.4,3.05)(6.3,2.55)(7.2,3.05)(7.2,4.05)(6.3,4.55)
\pspolygon[linewidth=0.05,fillstyle=solid,fillcolor=color422b](2.7,4.55)(1.8,4.05)(1.8,3.05)(2.7,2.55)(3.6,3.05)(3.6,4.05)(2.7,4.55)
\pspolygon[linewidth=0.05,fillstyle=solid,fillcolor=color422b](3.6,6.05)(2.7,5.55)(2.7,4.55)(3.6,4.05)(4.5,4.55)(4.5,5.55)(3.6,6.05)
\pspolygon[linewidth=0.05,fillstyle=solid,fillcolor=color422b](3.6,3.05)(2.7,2.55)(2.7,1.55)(3.6,1.05)(4.5,1.55)(4.5,2.55)(3.6,3.05)
\pspolygon[linewidth=0.05,fillstyle=solid,fillcolor=color422b](4.5,4.55)(3.6,4.05)(3.6,3.05)(4.5,2.55)(5.4,3.05)(5.4,4.05)(4.5,4.55)
\pspolygon[linewidth=0.05,fillstyle=solid,fillcolor=color422b](1.8,6.05)(0.9,5.55)(0.9,4.55)(1.8,4.05)(2.7,4.55)(2.7,5.55)(1.8,6.05)
\pspolygon[linewidth=0.05,fillstyle=solid,fillcolor=color422b](0.9,4.55)(0.0,4.05)(0.0,3.05)(0.9,2.55)(1.8,3.05)(1.8,4.05)(0.9,4.55)
\pspolygon[linewidth=0.05,fillstyle=solid,fillcolor=color422b](1.8,3.05)(0.9,2.55)(0.9,1.55)(1.8,1.05)(2.7,1.55)(2.7,2.55)(1.8,3.05)
\pspolygon[linewidth=0.05,fillstyle=solid,fillcolor=color422b](6.3,1.55)(5.4,1.05)(5.4,0.05)(6.3,-0.45)(7.2,0.05)(7.2,1.05)(6.3,1.55)
\pspolygon[linewidth=0.05,fillstyle=solid,fillcolor=color422b](7.2,3.05)(6.3,2.55)(6.3,1.55)(7.2,1.05)(8.1,1.55)(8.1,2.55)(7.2,3.05)
\pspolygon[linewidth=0.05,fillstyle=solid,fillcolor=color422b](7.2,0.05)(6.3,-0.45)(6.3,-1.45)(7.2,-1.95)(8.1,-1.45)(8.1,-0.45)(7.2,0.05)
\pspolygon[linewidth=0.05,fillstyle=solid,fillcolor=color422b](8.1,1.55)(7.2,1.05)(7.2,0.05)(8.1,-0.45)(9.0,0.05)(9.0,1.05)(8.1,1.55)
\pspolygon[linewidth=0.05,fillstyle=solid,fillcolor=color422b](5.4,3.05)(4.5,2.55)(4.5,1.55)(5.4,1.05)(6.3,1.55)(6.3,2.55)(5.4,3.05)
\pspolygon[linewidth=0.05,fillstyle=solid,fillcolor=color422b](4.5,1.55)(3.6,1.05)(3.6,0.05)(4.5,-0.45)(5.4,0.05)(5.4,1.05)(4.5,1.55)
\pspolygon[linewidth=0.05,fillstyle=solid,fillcolor=color422b](5.4,0.05)(4.5,-0.45)(4.5,-1.45)(5.4,-1.95)(6.3,-1.45)(6.3,-0.45)(5.4,0.05)
\pspolygon[linewidth=0.05,fillstyle=solid,fillcolor=color422b](10.8,3.05)(9.9,2.55)(9.9,1.55)(10.8,1.05)(11.7,1.55)(11.7,2.55)(10.8,3.05)
\pspolygon[linewidth=0.05,fillstyle=solid,fillcolor=color422b](11.7,4.55)(10.8,4.05)(10.8,3.05)(11.7,2.55)(12.6,3.05)(12.6,4.05)(11.7,4.55)
\pspolygon[linewidth=0.05,fillstyle=solid,fillcolor=color422b](11.7,1.55)(10.8,1.05)(10.8,0.05)(11.7,-0.45)(12.6,0.05)(12.6,1.05)(11.7,1.55)
\pspolygon[linewidth=0.05,fillstyle=solid,fillcolor=color422b](12.6,3.05)(11.7,2.55)(11.7,1.55)(12.6,1.05)(13.5,1.55)(13.5,2.55)(12.6,3.05)
\pspolygon[linewidth=0.05,fillstyle=solid,fillcolor=color422b](9.9,4.55)(9.0,4.05)(9.0,3.05)(9.9,2.55)(10.8,3.05)(10.8,4.05)(9.9,4.55)
\pspolygon[linewidth=0.05,fillstyle=solid,fillcolor=color422b](9.0,3.05)(8.1,2.55)(8.1,1.55)(9.0,1.05)(9.9,1.55)(9.9,2.55)(9.0,3.05)
\pspolygon[linewidth=0.05,fillstyle=solid,fillcolor=color422b](9.9,1.55)(9.0,1.05)(9.0,0.05)(9.9,-0.45)(10.8,0.05)(10.8,1.05)(9.9,1.55)
\pspolygon[linewidth=0.05,fillstyle=solid,fillcolor=color422b](11.7,7.55)(10.8,7.05)(10.8,6.05)(11.7,5.55)(12.6,6.05)(12.6,7.05)(11.7,7.55)
\pspolygon[linewidth=0.05,fillstyle=solid,fillcolor=color422b](12.6,9.05)(11.7,8.55)(11.7,7.55)(12.6,7.05)(13.5,7.55)(13.5,8.55)(12.6,9.05)
\pspolygon[linewidth=0.05,fillstyle=solid,fillcolor=color422b](12.6,6.05)(11.7,5.55)(11.7,4.55)(12.6,4.05)(13.5,4.55)(13.5,5.55)(12.6,6.05)
\pspolygon[linewidth=0.05,fillstyle=solid,fillcolor=color422b](13.5,7.55)(12.6,7.05)(12.6,6.05)(13.5,5.55)(14.4,6.05)(14.4,7.05)(13.5,7.55)
\pspolygon[linewidth=0.05,fillstyle=solid,fillcolor=color422b](10.8,9.05)(9.9,8.55)(9.9,7.55)(10.8,7.05)(11.7,7.55)(11.7,8.55)(10.8,9.05)
\pspolygon[linewidth=0.05,fillstyle=solid,fillcolor=color422b](9.9,7.55)(9.0,7.05)(9.0,6.05)(9.9,5.55)(10.8,6.05)(10.8,7.05)(9.9,7.55)
\pspolygon[linewidth=0.05,fillstyle=solid,fillcolor=color422b](10.8,6.05)(9.9,5.55)(9.9,4.55)(10.8,4.05)(11.7,4.55)(11.7,5.55)(10.8,6.05)
\pspolygon[linewidth=0.05,fillstyle=solid,fillcolor=color422b](8.1,10.55)(7.2,10.05)(7.2,9.05)(8.1,8.55)(9.0,9.05)(9.0,10.05)(8.1,10.55)
\pspolygon[linewidth=0.05,fillstyle=solid,fillcolor=color422b](9.0,12.05)(8.1,11.55)(8.1,10.55)(9.0,10.05)(9.9,10.55)(9.9,11.55)(9.0,12.05)
\pspolygon[linewidth=0.05,fillstyle=solid,fillcolor=color422b](9.0,9.05)(8.1,8.55)(8.1,7.55)(9.0,7.05)(9.9,7.55)(9.9,8.55)(9.0,9.05)
\pspolygon[linewidth=0.05,fillstyle=solid,fillcolor=color422b](9.9,10.55)(9.0,10.05)(9.0,9.05)(9.9,8.55)(10.8,9.05)(10.8,10.05)(9.9,10.55)
\pspolygon[linewidth=0.05,fillstyle=solid,fillcolor=color422b](7.2,12.05)(6.3,11.55)(6.3,10.55)(7.2,10.05)(8.1,10.55)(8.1,11.55)(7.2,12.05)
\pspolygon[linewidth=0.05,fillstyle=solid,fillcolor=color422b](6.3,10.55)(5.4,10.05)(5.4,9.05)(6.3,8.55)(7.2,9.05)(7.2,10.05)(6.3,10.55)
\pspolygon[linewidth=0.05,fillstyle=solid,fillcolor=color422b](7.2,9.05)(6.3,8.55)(6.3,7.55)(7.2,7.05)(8.1,7.55)(8.1,8.55)(7.2,9.05)
\pspolygon[linewidth=0.04,fillstyle=solid,fillcolor=color422b](10.0,-2.55)(9.1,-3.05)(9.1,-4.05)(10.0,-4.55)(10.9,-4.05)(10.9,-3.05)(10.0,-2.55)
\pspolygon[linewidth=0.14](6.3,7.55)(5.4,7.05)(5.4,6.05)(4.5,5.55)(4.5,4.55)(5.4,4.05)(5.4,3.05)(6.3,2.55)(7.2,3.05)(8.1,2.55)(9.0,3.05)(9.0,4.05)(9.9,4.55)(9.9,5.55)(9.0,6.05)(9.0,7.05)(8.1,7.55)(7.2,7.05)(6.3,7.55)
\pspolygon[linewidth=0.04,fillstyle=solid,fillcolor=color422b](14.5,-10.05)(13.6,-10.55)(13.6,-11.55)(14.5,-12.05)(15.4,-11.55)(15.4,-10.55)(14.5,-10.05)
\pspolygon[linewidth=0.04,fillstyle=solid,fillcolor=color422b](11.8,-2.55)(10.9,-3.05)(10.9,-4.05)(11.8,-4.55)(12.7,-4.05)(12.7,-3.05)(11.8,-2.55)
\pspolygon[linewidth=0.04,fillstyle=solid,fillcolor=color422b](12.7,-4.05)(11.8,-4.55)(11.8,-5.55)(12.7,-6.05)(13.6,-5.55)(13.6,-4.55)(12.7,-4.05)
\pspolygon[linewidth=0.04,fillstyle=solid,fillcolor=color422b](10.9,-4.05)(10.0,-4.55)(10.0,-5.55)(10.9,-6.05)(11.8,-5.55)(11.8,-4.55)(10.9,-4.05)
\pspolygon[linewidth=0.04,fillstyle=solid,fillcolor=color422b](9.1,-4.05)(8.2,-4.55)(8.2,-5.55)(9.1,-6.05)(10.0,-5.55)(10.0,-4.55)(9.1,-4.05)
\pspolygon[linewidth=0.04,fillstyle=solid,fillcolor=color422b](10.0,-5.55)(9.1,-6.05)(9.1,-7.05)(10.0,-7.55)(10.9,-7.05)(10.9,-6.05)(10.0,-5.55)
\pspolygon[linewidth=0.04,fillstyle=solid,fillcolor=color422b](8.2,-5.55)(7.3,-6.05)(7.3,-7.05)(8.2,-7.55)(9.1,-7.05)(9.1,-6.05)(8.2,-5.55)
\pspolygon[linewidth=0.04,fillstyle=solid,fillcolor=OliveGreen](11.8,-5.55)(10.9,-6.05)(10.9,-7.05)(11.8,-7.55)(12.7,-7.05)(12.7,-6.05)(11.8,-5.55)
\pspolygon[linewidth=0.04,fillstyle=solid,fillcolor=OliveGreen](12.7,-7.05)(11.8,-7.55)(11.8,-8.55)(12.7,-9.05)(13.6,-8.55)(13.6,-7.55)(12.7,-7.05)
\pspolygon[linewidth=0.04,fillstyle=solid,fillcolor=OliveGreen](10.9,-7.05)(10.0,-7.55)(10.0,-8.55)(10.9,-9.05)(11.8,-8.55)(11.8,-7.55)(10.9,-7.05)
\pspolygon[linewidth=0.16](11.8,-5.55)(10.9,-6.05)(10.9,-7.05)(10.0,-7.55)(10.0,-8.55)(10.9,-9.05)(11.8,-8.55)(12.7,-9.05)(13.6,-8.55)(13.6,-7.55)(12.7,-7.05)(12.7,-6.05)(11.8,-5.55)
\pspolygon[linewidth=0.04,fillstyle=solid,fillcolor=color422b](16.3,-7.05)(15.4,-7.55)(15.4,-8.55)(16.3,-9.05)(17.2,-8.55)(17.2,-7.55)(16.3,-7.05)
\pspolygon[linewidth=0.04,fillstyle=solid,fillcolor=color422b](7.3,-7.05)(6.4,-7.55)(6.4,-8.55)(7.3,-9.05)(8.2,-8.55)(8.2,-7.55)(7.3,-7.05)
\pspolygon[linewidth=0.04,fillstyle=solid,fillcolor=color422b](9.1,-7.05)(8.2,-7.55)(8.2,-8.55)(9.1,-9.05)(10.0,-8.55)(10.0,-7.55)(9.1,-7.05)
\pspolygon[linewidth=0.04,fillstyle=solid,fillcolor=color422b](10.0,-8.55)(9.1,-9.05)(9.1,-10.05)(10.0,-10.55)(10.9,-10.05)(10.9,-9.05)(10.0,-8.55)
\pspolygon[linewidth=0.04,fillstyle=solid,fillcolor=color422b](8.2,-8.55)(7.3,-9.05)(7.3,-10.05)(8.2,-10.55)(9.1,-10.05)(9.1,-9.05)(8.2,-8.55)
\pspolygon[linewidth=0.04,fillstyle=solid,fillcolor=color422b](14.5,-4.05)(13.6,-4.55)(13.6,-5.55)(14.5,-6.05)(15.4,-5.55)(15.4,-4.55)(14.5,-4.05)
\pspolygon[linewidth=0.04,fillstyle=solid,fillcolor=color422b](15.4,-5.55)(14.5,-6.05)(14.5,-7.05)(15.4,-7.55)(16.3,-7.05)(16.3,-6.05)(15.4,-5.55)
\pspolygon[linewidth=0.04,fillstyle=solid,fillcolor=color422b](13.6,-5.55)(12.7,-6.05)(12.7,-7.05)(13.6,-7.55)(14.5,-7.05)(14.5,-6.05)(13.6,-5.55)
\pspolygon[linewidth=0.04,fillstyle=solid,fillcolor=color422b](11.8,-8.55)(10.9,-9.05)(10.9,-10.05)(11.8,-10.55)(12.7,-10.05)(12.7,-9.05)(11.8,-8.55)
\pspolygon[linewidth=0.04,fillstyle=solid,fillcolor=color422b](12.7,-10.05)(11.8,-10.55)(11.8,-11.55)(12.7,-12.05)(13.6,-11.55)(13.6,-10.55)(12.7,-10.05)
\pspolygon[linewidth=0.04,fillstyle=solid,fillcolor=color422b](10.9,-10.05)(10.0,-10.55)(10.0,-11.55)(10.9,-12.05)(11.8,-11.55)(11.8,-10.55)(10.9,-10.05)
\pspolygon[linewidth=0.04](9.1,-10.05)(8.2,-10.55)(8.2,-11.55)(9.1,-12.05)(10.0,-11.55)(10.0,-10.55)(9.1,-10.05)
\pspolygon[linewidth=0.04,fillstyle=solid,fillcolor=color422b](9.1,-10.05)(8.2,-10.55)(8.2,-11.55)(9.1,-12.05)(10.0,-11.55)(10.0,-10.55)(9.1,-10.05)
\pspolygon[linewidth=0.04,fillstyle=solid,fillcolor=color422b](13.6,-2.55)(12.7,-3.05)(12.7,-4.05)(13.6,-4.55)(14.5,-4.05)(14.5,-3.05)(13.6,-2.55)
\pspolygon[linewidth=0.04,fillstyle=solid,fillcolor=color422b](14.5,-7.05)(13.6,-7.55)(13.6,-8.55)(14.5,-9.05)(15.4,-8.55)(15.4,-7.55)(14.5,-7.05)
\pspolygon[linewidth=0.04,fillstyle=solid,fillcolor=color422b](15.4,-8.55)(14.5,-9.05)(14.5,-10.05)(15.4,-10.55)(16.3,-10.05)(16.3,-9.05)(15.4,-8.55)
\pspolygon[linewidth=0.04,fillstyle=solid,fillcolor=color422b](13.6,-8.55)(12.7,-9.05)(12.7,-10.05)(13.6,-10.55)(14.5,-10.05)(14.5,-9.05)(13.6,-8.55)
\pspolygon[linewidth=0.16](11.8,-2.55)(10.9,-3.05)(10.9,-4.05)(10.0,-4.55)(10.0,-5.55)(10.9,-6.05)(11.8,-5.55)(12.7,-6.05)(13.6,-5.55)(13.6,-4.55)(12.7,-4.05)(12.7,-3.05)(11.8,-2.55)
\pspolygon[linewidth=0.16](9.1,-4.05)(8.2,-4.55)(8.2,-5.55)(7.3,-6.05)(7.3,-7.05)(8.2,-7.55)(9.1,-7.05)(10.0,-7.55)(10.9,-7.05)(10.9,-6.05)(10.0,-5.55)(10.0,-4.55)(9.1,-4.05)
\pspolygon[linewidth=0.16](11.8,-8.55)(10.9,-9.05)(10.9,-10.05)(10.0,-10.55)(10.0,-11.55)(10.9,-12.05)(11.8,-11.55)(12.7,-12.05)(13.6,-11.55)(13.6,-10.55)(12.7,-10.05)(12.7,-9.05)(11.8,-8.55)
\pspolygon[linewidth=0.16](9.1,-7.05)(8.2,-7.55)(8.2,-8.55)(7.3,-9.05)(7.3,-10.05)(8.2,-10.55)(9.1,-10.05)(10.0,-10.55)(10.9,-10.05)(10.9,-9.05)(10.0,-8.55)(10.0,-7.55)(9.1,-7.05)
\pspolygon[linewidth=0.16](14.5,-7.05)(13.6,-7.55)(13.6,-8.55)(12.7,-9.05)(12.7,-10.05)(13.6,-10.55)(14.5,-10.05)(15.4,-10.55)(16.3,-10.05)(16.3,-9.05)(15.4,-8.55)(15.4,-7.55)(14.5,-7.05)
\pspolygon[linewidth=0.16](14.5,-4.05)(13.6,-4.55)(13.6,-5.55)(12.7,-6.05)(12.7,-7.05)(13.6,-7.55)(14.5,-7.05)(15.4,-7.55)(16.3,-7.05)(16.3,-6.05)(15.4,-5.55)(15.4,-4.55)(14.5,-4.05)
\pspolygon[linewidth=0.14](9.9,4.55)(9.0,4.05)(9.0,3.05)(8.1,2.55)(8.1,1.55)(9.0,1.05)(9.0,0.05)(9.9,-0.45)(10.8,0.05)(11.7,-0.45)(12.6,0.05)(12.6,1.05)(13.5,1.55)(13.5,2.55)(12.6,3.05)(12.6,4.05)(11.7,4.55)(10.8,4.05)(9.9,4.55)
\pspolygon[linewidth=0.14](10.8,9.05)(9.9,8.55)(9.9,7.55)(9.0,7.05)(9.0,6.05)(9.9,5.55)(9.9,4.55)(10.8,4.05)(11.7,4.55)(12.6,4.05)(13.5,4.55)(13.5,5.55)(14.4,6.05)(14.4,7.05)(13.5,7.55)(13.5,8.55)(12.6,9.05)(11.7,8.55)(10.8,9.05)
\pspolygon[linewidth=0.14](7.2,12.05)(6.3,11.55)(6.3,10.55)(5.4,10.05)(5.4,9.05)(6.3,8.55)(6.3,7.55)(7.2,7.05)(8.1,7.55)(9.0,7.05)(9.9,7.55)(9.9,8.55)(10.8,9.05)(10.8,10.05)(9.9,10.55)(9.9,11.55)(9.0,12.05)(8.1,11.55)(7.2,12.05)
\pspolygon[linewidth=0.14](2.7,10.55)(1.8,10.05)(1.8,9.05)(0.9,8.55)(0.9,7.55)(1.8,7.05)(1.8,6.05)(2.7,5.55)(3.6,6.05)(4.5,5.55)(5.4,6.05)(5.4,7.05)(6.3,7.55)(6.3,8.55)(5.4,9.05)(5.4,10.05)(4.5,10.55)(3.6,10.05)(2.7,10.55)
\pspolygon[linewidth=0.14](1.8,6.05)(0.9,5.55)(0.9,4.55)(0.0,4.05)(0.0,3.05)(0.9,2.55)(0.9,1.55)(1.8,1.05)(2.7,1.55)(3.6,1.05)(4.5,1.55)(4.5,2.55)(5.4,3.05)(5.4,4.05)(4.5,4.55)(4.5,5.55)(3.6,6.05)(2.7,5.55)(1.8,6.05)
\pspolygon[linewidth=0.14](5.4,3.05)(4.5,2.55)(4.5,1.55)(3.6,1.05)(3.6,0.05)(4.5,-0.45)(4.5,-1.45)(5.4,-1.95)(6.3,-1.45)(7.2,-1.95)(8.1,-1.45)(8.1,-0.45)(9.0,0.05)(9.0,1.05)(8.1,1.55)(8.1,2.55)(7.2,3.05)(6.3,2.55)(5.4,3.05)
\end{pspicture}
}
\caption{The cellular layout of $B=3$ and $B=7$ clustered network MIMO coordination. The borders of clusters are bold. Green colored cells represent the analyzed center cluster and the grey cells are causing inter-cell interference. For $B=7$, one tier of interfering clusters is considered, while for $B=3$ two tiers of interfering cells are accounted for.}
\label{fig:CellularLayout}
\end{figure}
\par Our cellular network setup involves clustering. Since global coordination is not feasible, clustering with cluster sizes of up to $B=7$ is considered. The cellular network layout is shown in Fig.~\ref{fig:CellularLayout}. A base station is located at the center of each hexagonal cell. Each base station is equipped with $n_t$ transmit antennas. There are $n_r$ receive antennas on each user's receiver and there are $K$ users per cell per subband. All $N_t=Bn_t$ base stations' transmit antennas in each cluster are coordinated. In Fig.~\ref{fig:CellularLayout} the clusters of sizes 3 and 7 are shown. For cluster size 7, one wrap-around layer of clusters is considered to contribute inter-cluster interference, while for $B=3$ two tiers of interfering cells are accounted for. User locations are generated randomly, uniformly and independently in each cell. For each drop of users, the distance of users from base stations in the network is computed and path loss, lognormal and Rayleigh fading are included in the channel gain calculations. User scheduling is performed employing a greedy algorithm with maximum sum rate and proportionally-fair criteria with the updated weights for the rate of each user as in \cite{Viswanathan03,Sigdel08,Sigdel09}. To compare the results all the sum rates achieved through network MIMO coordination are normalized by the size of clusters $B$. Base stations causing inter-cluster interference are assumed to transmit at full power, which is the worst case as discussed in Section~\ref{Sec:SystemModel}.

\section{Numerical Results}\label{Sec:NumericalResults}
In this section, the performance results (obtained via Monte Carlo simulations) of the proposed optimal BD scheme in a network MIMO coordinated system are discussed. The network MIMO coordination exhibits several system advantages, which are exposed in the following.

\begin{figure}[!t]
\centering
 \includegraphics[scale=.4]{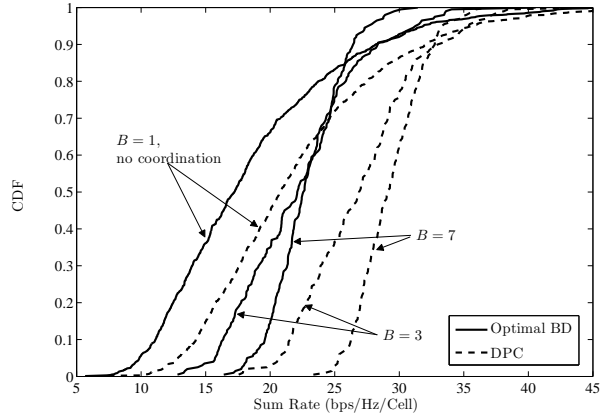}
\caption{CDF of sum rate with different cluster sizes $B=1,3,7$, $n_t=4$, $n_r=2$ and 10 users per cell.}
\label{fig:NetworkMIMO}\end{figure}
\subsection{Network MIMO Gains}
While the universal network MIMO coordination is practically impossible, clustering is a practical scheme, which also benefits the network MIMO coordination gains and reduces the amount of feedback required at the base stations \cite{Zhang09,Huang09}. The size of clusters, $B$, is a parameter in network MIMO coordination. $B=1$ means no coordination with optimal BD scheme applied. Fig.~\ref{fig:NetworkMIMO} shows that with increasing cluster size throughput of the system increases. System throughput is computed using MSR scheduling and averaged over several channel realizations for a large number of user locations generated randomly. The normalized throughput for different cluster sizes is compared, which means that the total throughput in each cluster is divided by the number of cells in each cluster $B$. The normalized sum rate has lower variance in larger clusters, which shows that the performance of the system is less dependent on the position of users and that network MIMO coordination brings more stability to the system.

\begin{figure}[!t]
\centering
 \includegraphics[scale=.4]{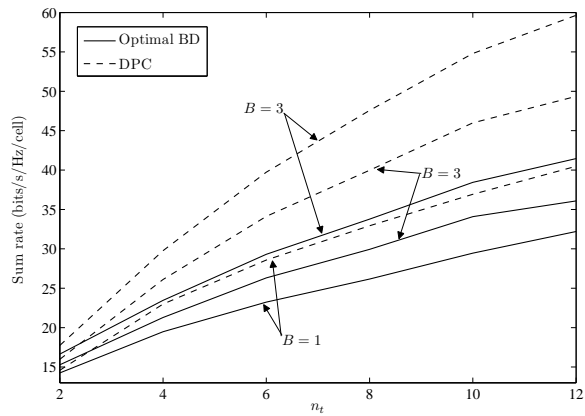}
\caption{Sum-rate increase with the number of antennas per base station. $n_r=2$.}
\label{fig:MultipleAntenna}
\end{figure}

\subsection{Multiple-Antenna Gains}
The inter-cell interference mitigation through coordination of base stations enables the cellular network to enjoy the great spectral efficiency improvement associated with employing multiple antennas. Fig.~\ref{fig:MultipleAntenna} shows the linear growth of the maximum throughput achievable through the proposed optimal multi-cell BD and the capacity limits of DPC \cite{Yu07}. The number of receive antennas at each mobile user is fixed to $n_r=2$ and the number of transmit antennas $n_t$ at each base station is increasing. When the cluster size grows, the slope of spectral efficiency also increases. The maximum power on each transmit antenna is normalized such that total power at each base station for different $n_t$ is constant.
\begin{figure}[!t]
\centering
 \includegraphics[scale=.4]{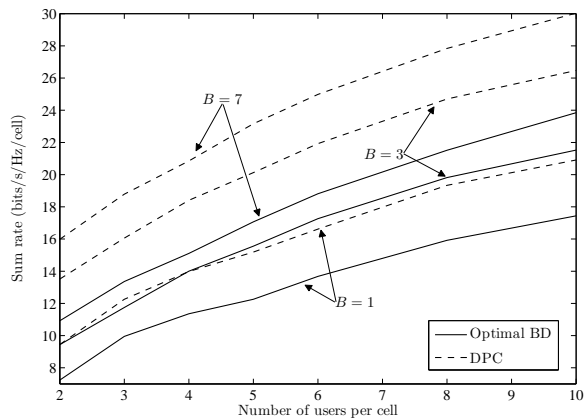}
\caption{Sum rate per cell achieved with the proposed optimal BD and the capacity limits of DPC for cluster sizes $B=1,3,7$; $n_t=4$, $n_r=2$.}
\label{fig:MultiuserDiversity1}
\end{figure}

\subsection{Multiuser Diversity}
Multi-cell coordination benefits from increased multiuser diversity, since the number of users scheduled at each time interval is $B$ times of that without coordination. In Fig.~\ref{fig:MultiuserDiversity1}, the multiuser diversity gain of network MIMO is shown with up to 10 users per cell. The MSR scheduling is applied for each drop of users and averaged over several channel realizations.

\begin{figure}[!t]
\centering
 \includegraphics[scale=.4]{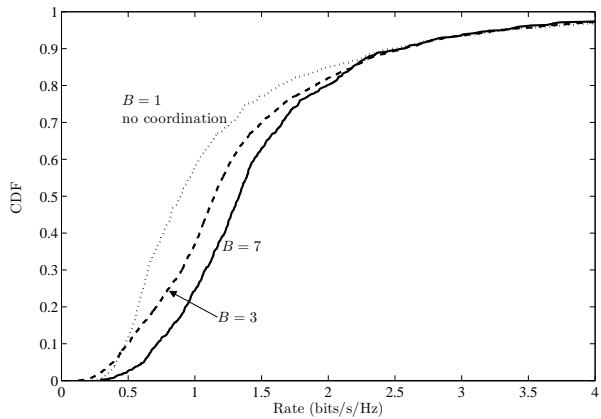}
\caption{CDF of the mean rates in the clusters of sizes $B=3,7$ and comparison with $B=1$ (no coordination) using the proposed optimal BD.}
\label{fig:Fairness}\end{figure}

\subsection{Fairness Advantages}
One of the main purposes of network MIMO coordination is that the cell-edge users gain from neighboring base stations signals. In Fig.~\ref{fig:Fairness}, the cumulative distribution functions (CDFs) of mean rates for users are shown and compared for $B=1$ (i.e. beamforming without coordination) and $B=3,7$ for the proposed optimal BD scheme. There are 10 users per cell randomly and uniformly dropped in the network for each simulation. For each drop of users, the proportionally fair scheduling algorithm is applied over hundreds of scheduling time intervals using sliding window width $\tau=10$ time slots (see \cite{Vishwanath03}). Each user's rates achieved in all time intervals are averaged to find the mean rates per user and their CDF for several user locations is plotted. As shown by the plots, for $B=3$ and $B=7$ network MIMO coordination nearly 70\% and 80\% users have mean rate larger than 1 bps/Hz, respectively, while for the scheme without coordination it is 45\% of users. However, fairness among users does not seem to be improved when cluster sizes increases. This is perhaps due to the existence of larger number of cell-edge users when cluster size increases.
\begin{figure}[!t]
\centering
 \includegraphics[scale=.4]{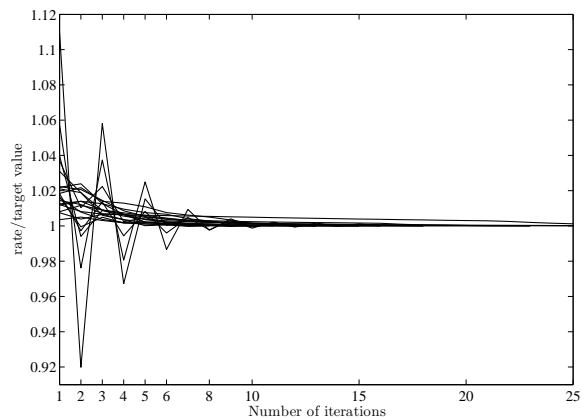}
\caption{Convergence of the gradient descent method for the proposed optimal BD for $B=3$, $n_t=4$, $n_r=2$, and 8 users per cell.}
\label{fig:Iterations}
\end{figure}
\subsection{Convergence}
Convergence of the gradient descent method proposed in Section \ref{Sec:OptimalBD} is illustrated in Fig.~\ref{fig:Iterations}. The normalized sum rates obtained after each iteration with respect to the optimal target values versus the number of iterations are depicted. The convergence behavior of the algorithm for 20 independent and randomly generated user location sets is shown, and their channel realizations are tested with the proposed iterative algorithm and the values of sum rate after each iteration divided by the target value are monitored. Nearly all of the optimizations converge to the target value within only 10 first iterations with 1\% error.


\section{Conclusions}\label{Sec:Conclusions}
In this paper, a multi-cell coordinated downlink MIMO transmission has been considered under per-antenna power constraints. Sub-optimality of the conventional BD considered in earlier research has been shown and it has motivated the search for the optimal BD scheme. The optimal block diagonalization (BD) scheme for network MIMO coordinated system under per-antenna power constraints has been proposed in this paper and it has been shown that it can be generalized to the case of per-base station power constraints. The simple iterative descent gradient algorithm employed in this paper gives the optimal precoders for multi-cell BD. The comprehensive simulation results have demonstrated advantages achieved by using multi-cell coordinated transmission under more practical per-antenna power constraints.
\appendix[Proof of Theorem 1]\label{App:A}
We consider the optimization problem (\ref{optimization1}). For the ease of further analysis, let us substitute $\mathbf S_k=\mathbf H_k\mathbf V_k\mathbf \Phi_k \mathbf V^\textsf{H}_k\mathbf H_k^\textsf{H}$ and $\mathbf G_k=\mathbf H_k\mathbf V_k$, where $\rank(\mathbf G_k)\leq n_r$. Note that the rank constraint on $\mathbf \Phi_k$ must be inserted into the optimization when $m_r > n_r$, and hence it makes the problem non-convex. Thus, to analyze this problem two cases are considered based on the value of $m_r$ with respect to $n_r$. In the first case $m_r=n_r$, when the total number of transmit antennas at all base stations, $N_t$, is equal to the total number of receive antennas at all $K$ served users, $N_r$. In the second case $N_t > N_r$.
 \subsection{$N_t=N_r$}\label{m_r=n_r}
 This happens when exactly $K=\frac{N_t}{n_r}$ users are scheduled. In this case, the rank constraint over $\mathbf \Phi_k$ can be dropped because $m_r=n_r$ and therefore the optimization problem in (\ref{optimization1}) is convex. The matrices $\mathbf G_k$ are also square and invertible. Therefore $\mathbf G_k^\dag=\mathbf G_k^{-1}$. Let $\mathbf Q_k=\mathbf V_k\mathbf G_k^{-1}$ which is an $N_t \times n_r$ matrix.  Thus, the throughput maximization problem can be expressed as (since $\mathbf S_k \succeq 0 \Longleftrightarrow \mathbf G_k^{-1}\mathbf S_k \mathbf G_k^{-\textsf{H}}$)
\begin{equation}\label{optimization2}
\begin{array}{rl}
\text{maximize}& \sum_{k=1}^K\log\left|\mathbf I+\mathbf S_k\right|\\
 \text{subject to} & \left[\sum_{k=1}^K\mathbf Q_k\mathbf S_k \mathbf Q^\textsf{H}_k\right]_{i,i} \leq P_i,\quad  i=1,\ldots,N_t\\
&  \mathbf S_k \succeq 0, \quad k=1,\ldots,K,
\end{array}
\end{equation}
where $\mathbf S_k \in \mathbb C^{n_r \times n_r}$. Although one possibility is to perform this convex optimization with $Kn_r(n_r-1)/2$ variables introducing logarithmic barrier functions for inequality power constraints and the set of positive semi-definite constraints, we approach the problem by establishing the dual problem and solving it through simple and efficient gradient descent method \cite{Boyd04}. Hence, the Lagrangian function can be formed as
\begin{equation}\label{Lagrangian2}
\begin{split}
\mathcal L( \left\{\mathbf S\right\}; \mathbf \Lambda)=&\sum_{k=1}^K \log \left|\mathbf I + \mathbf S_k\right|+\sum_{k=1}^K\tr\left\{\mathbf \Omega_k\mathbf S_k\right\}\\&-\tr\left\{\mathbf \Lambda\left(\sum_{k=1}^K\mathbf Q_k\mathbf S_k\mathbf Q_k^\textsf{H}-\mathbf P\right)\right\}
\end{split}
\end{equation}
where $\mathbf \Lambda=\diag(\lambda_1,\ldots,\lambda_{N_t})$ is a dual variable which is a diagonal matrix with non-negative elements, $\lambda_i \geq 0$. The positive semi-definite matrix $\mathbf \Omega_k$ is a dual variable to assure positive semi-definiteness of $\mathbf S_k$. The Karush-Kuhn-Tucker (KKT) conditions require that the optimal values of primal and dual variables \cite{Boyd04} satisfy the following
\begin{equation}\label{eqn:KKTConditions}
\begin{split}
\mathbf S_k&=\left(\mathbf Q_k^\textsf{H}\mathbf \Lambda \mathbf Q_k-\mathbf \Omega_k\right)^{-1}-\mathbf I,\\
\mathbf S_k & \succeq 0,\\
\tr\left\{\mathbf \Omega_k\mathbf S_k\right\}&=0,\mathbf \Omega_k \succeq 0\\
\tr\left\{\mathbf \Lambda\left(\sum_{k=1}^K\mathbf Q_k\mathbf S_k\mathbf Q_k^\textsf{H}-\mathbf P\right)\right\}&=0,\mathbf \Lambda \succeq 0\\
\mathbf P & \succeq \diag\left[\sum_{k=1}^K\mathbf Q_k\mathbf S_k\mathbf Q_k^\textsf{H}\right] .
\end{split}
\end{equation}
Let the SVD of $\mathbf Q_k^\textsf{H} \mathbf \Lambda \mathbf Q_k=\mathbf U_{k}\mathbf \Sigma_k \mathbf U_k^\textsf{H}$. Since $\mathbf Q_k^\textsf{H} \mathbf \Lambda \mathbf Q_k \succeq 0$, the diagonal entries of $\mathbf \Sigma_k$ are the eigenvalues of $\mathbf Q_k^\textsf{H} \mathbf \Lambda \mathbf Q_k$. The first KKT condition on $\mathbf S_k$ and $\mathbf \Omega_k$ requires that
\begin{equation}\label{Omega_k}
\mathbf \Omega_k=\mathbf U_k\left[\mathbf \Sigma_k-\mathbf I\right]_+\mathbf U_k^\textsf{H},
\end{equation}
where the operator $\left[\mathbf D\right]_+=\diag\left[\max(0,d_{1}),\ldots,\max(0,d_n)\right]$ on a diagonal matrix $\mathbf D=\diag\left[d_1,\ldots,d_n\right]$.
Replacing these in KKT condition corresponding to the power constraints gives
\begin{equation}
\begin{split}
\tr\left\{\!\mathbf \Lambda\!\left(\!\sum_{k=1}^K\mathbf Q_k\mathbf S_k\mathbf Q_k^\textsf{H}\!-\!\mathbf P\!\right)\!\right\}=&Kn_r-\tr\left\{\mathbf \Lambda \mathbf P\right\}\\
&-\tr\!\left\{\sum_{k=1}^K \!\left(\mathbf Q_k^\textsf{H} \mathbf \Lambda \mathbf Q_k\!-\mathbf \Omega_k\right)\!\right\}.
\end{split}
\end{equation}
Now, we establish the Lagrange dual function as
\begin{equation}\label{gFunction}
\begin{split}
g\left(\mathbf \Lambda\right)=&\sup_{\mathbf S_k}\mathcal L\left(\left\{\mathbf S\!\right\}\!\right)\\
=& -\sum_{k=1}^K \log\left|\mathbf Q_k^\textsf{H}\mathbf \Lambda \mathbf Q_k -\mathbf \Omega_k\right|-Kn_r\\
& +\tr\left\{\sum_{k=1}^K\left(\mathbf Q_k^\textsf{H}\mathbf \Lambda \mathbf Q_k -\mathbf \Omega_k\right)\right\}+\tr\left\{\mathbf \Lambda\mathbf P\right\}.
\end{split}
\end{equation}
Since the constraint functions are affine, strong duality holds and thus the dual objective reaches a minimum at the optimal value of the primal problem \cite{Boyd04}. As a result, the Lagrange dual problem can be stated as
\begin{equation}\label{DualProblem}
\begin{array}{cl}
\text{minimize}& g(\mathbf \Lambda)\\
 \text{subject to} & \mathbf \Lambda \succeq 0, \quad \mathbf \Lambda \text{ diagonal}
\end{array}
\end{equation}
The gradient of $g$ can be obtained from (\ref{gFunction}) as
\begin{equation}\label{GradientDescentSearchDirection}
\begin{split}
\nabla_{\mathbf \Lambda}g=&-\sum_{k=1}^K \diag\left[\mathbf Q_k \left(\mathbf Q_k^\textsf{H} \mathbf \Lambda \mathbf Q_k-\mathbf \Omega_k\right)^{-1}\mathbf Q_k^\textsf{H}\right]\\&+\mathbf P+\sum_{k=1}^K \diag\left[\mathbf Q_k \mathbf Q_k^\textsf{H}\right].
\end{split}
\end{equation}
This gives a descent search direction, $\Delta \mathbf \Lambda=-\nabla_{\mathbf \Lambda}g$, for the gradient algorithm for the Lagrange dual problem \cite{Boyd04}.

 \subsection{$N_t > N_r$}\label{m_r>n_r}
When the total number of transmit antennas is strictly larger than the total number of receive antennas in the network (i.e. $N_t > N_r$) the optimization problem in (\ref{optimization2}) is no longer convex due to the rank constraints. We relax the problem and show that it leads to an optimal solution, which also satisfies the rank constraints in the original problem. Similar gradient algorithm to the one for $N_t=N_r$ can be deployed to find the optimal BD precoders.
\par Recall that $m_r=N_t-(K-1)n_r$. Thus, when the total number of transmit antennas is strictly larger than the total number of receive antennas, $N_t  > N_r$, then $m_r > n_r$. From Section \ref{Sec:Multi-CellBD} note that $\mathbf V_k$ is an $N_t \times m_r$ matrix and correspondingly the size of $\mathbf \Psi_k$ is $m_r \times n_r$ which enforces a rank constraint over $\mathbf \Phi_k=\mathbf \Psi_k \mathbf \Psi_k^\textsf{H}$. (i.e. $\rank(\mathbf \Phi_k) \leq n_r$). This updates the optimization in (\ref{optimization1}) by adding the rank constraints as
\begin{equation}\label{optimization3}
\begin{array}{cl}
\text{maximize}& \sum_{k=1}^K\log\left|\mathbf I+\mathbf H_k\mathbf V_k\mathbf \Phi_k \mathbf V^\textsf{H}_k\mathbf H_k^\textsf{H}\right|\\
 \text{subject to} & \left[\sum_{k=1}^K\mathbf V_k\mathbf \Phi_k \mathbf V^\textsf{H}_k\right]_{i,i} \leq p_i,\quad  i=1,\ldots,N_t\\
&  \mathbf \Phi_k \succeq 0, \quad \rank(\mathbf \Phi_k) \leq n_r \quad k=1,\ldots,K.
\end{array}
\end{equation}
The problem above is not convex due to the rank constraint. Assume the convex relaxation problem obtained by removing the rank constraint. The problem can then be expressed as
\begin{equation}\label{optimization4}
\begin{array}{rl}
\text{maximize}& \sum_{k=1}^K\log\left|\mathbf I+\mathbf H_k\mathbf V_k\mathbf \Phi_k \mathbf V^\textsf{H}_k\mathbf H_k^\textsf{H}\right|\\
 \text{subject to} & \left[\sum_{k=1}^K\mathbf V_k\mathbf \Phi_k \mathbf V^\textsf{H}_k\right]_{i,i} \leq p_i,\quad  i=1,\ldots,N_t\\
&  \mathbf \Phi_k \succeq 0, \quad k=1,\ldots,K
\end{array}
\end{equation}
Since this problem is convex and the constraints are affine, any solution satisfying the KKT conditions is optimal \cite{Boyd04}. Let us introduce an optimization problem
\begin{equation}\label{optimization5}
\begin{array}{rl}
\text{maximize}& \sum_{k=1}^K\log\left|\mathbf I+\mathbf S_k\right|\\
 \text{subject to} & \left[\sum_{k=1}^K\mathbf V_k\mathbf G_k^\dag  \mathbf S_k \left(\mathbf G_k^\dag\right)^\textsf{H}\mathbf V^\textsf{H}_k\right]_{i,i} \leq p_i,\\
 & \quad  i=1,\ldots,N_t\\
&  \mathbf S_k \succeq 0, \quad k=1,\ldots,K
\end{array}
\end{equation}
 Assume the optimal solutions for this problem are $\mathbf S^{\star}_k$s. Defining $\mathbf \Phi_k=\mathbf G_k^\dag \mathbf S_k^\star \left(\mathbf G_k^\dag\right)^\textsf{H}$, satisfies all the KKT conditions for (\ref{optimization4}), since $\mathbf G_k \mathbf \Phi_k \mathbf G_k^\textsf{H}=\mathbf S_k^\star$. Furthermore, $\rank(\mathbf \Phi_k)=\rank(\mathbf S_k)\leq n_r$ which also satisfies the rank constraint in the original optimization problem (\ref{optimization3}). Note that also $\mathbf \Phi_k \succeq 0  \Leftrightarrow \mathbf S_k^\star \succeq 0$ (see \cite[p. 399]{Horn85}).
\par The optimization in (\ref{optimization5}) is equivalent to the convex optimization problem in (\ref{optimization2}) by replacing $\mathbf Q_k=\mathbf V_k\mathbf G_k^\dag$. Recall that when $m_r=n_r$ then the matrix $\mathbf G_k$ is square and invertible. Hence, $\mathbf Q_k=\mathbf V_k \mathbf G_k^{-1}$, as defined in Section \ref{m_r=n_r}. Thus, this problem can be solved through the gradient descent method applied to the dual problem (\ref{DualProblem}) with the gradient descent search direction (\ref{GradientDescentSearchDirection}). The stopping criterion is also the same as (\ref{StoppingCriterion}) except that $\mathbf Q_k$ has different definition.
\par Note that (\ref{eqn:OptimalBDPrecoder}) can be simply concluded from the first equation of the KKT conditions (\ref{eqn:KKTConditions}) and the definition of $\mathbf \Phi_k=\mathbf G_k^\dag \mathbf S_k^\star \left(\mathbf G_k^\dag\right)^\mathsf H$ for the optimal value of dual variables $\mathbf \Lambda^\star$.
\section*{Acknowledgment}
Funding for this work has been provided by the Natural Sciences and Engineering Research Council (NSERC) of Canada,
TRLabs, the Rohit Sharma Professorship, Alberta Innovates – Technology Futures and the University of Alberta.
\bibliographystyle{IEEEtran}
\bibliography{IEEEabrv,MyBib}


\end{document}